\documentclass[11pt,preprint]{aastex}
\usepackage{graphicx}
\usepackage{natbib}

\newcommand{\kms}{\ensuremath{\rm km\,s^{-1}}}
\newcommand{\ms}{\ensuremath{\rm m\,s^{-1}}}

\newcommand{\gcmc}{\ensuremath{\rm g\,cm^{-3}}}

\newcommand{\rhk}{\ensuremath{R^{\prime}_{HK}}}
\newcommand{\logrhk}{\ensuremath{\log\rhk}}

\newcommand{\rstar}{\ensuremath{R_\star}}

\newcommand{\rpl}{\ensuremath{R_{p}}}
\newcommand{\mpl}{\ensuremath{M_{p}}}

\newcommand{\gpl}{\ensuremath{g_{p}}}

\newcommand{\rjup}{\ensuremath{R_{\rm Jup}}}
\newcommand{\mjup}{\ensuremath{M_{\rm Jup}}}

\shorttitle{Spectro-Photometry of TrES-3 and TrES-4}
\shortauthors{Sozzetti et al.}

\begin{document}


\title{A New Spectroscopic and Photometric Analysis of the Transiting
Planet Systems TrES-3 and TrES-4}


\author{Alessandro Sozzetti\altaffilmark{1,2}, 
Guillermo Torres\altaffilmark{1}, David
Charbonneau\altaffilmark{1,12}, Joshua N.\ Winn\altaffilmark{3}, 
Sylvain G. Korzennik\altaffilmark{1}, Matthew J.\ Holman\altaffilmark{1}, 
David W.\ Latham\altaffilmark{1}, John B.\ Laird\altaffilmark{4}, Jos\'e Fernandez\altaffilmark{1}, 
Francis T.\ O'Donovan\altaffilmark{5,6}, Georgi Mandushev\altaffilmark{7}, Edward Dunham\altaffilmark{7}, 
Mark E. Everett\altaffilmark{8}, Gilbert A. Esquerdo\altaffilmark{1}, Markus Rabus\altaffilmark{9}, 
Juan A. Belmonte\altaffilmark{9}, Hans J. Deeg\altaffilmark{9}, Timothy N. Brown\altaffilmark{10,11}, 
M\'arton G. Hidas\altaffilmark{10,11}, and Nairn Baliber\altaffilmark{10,11}} 
\altaffiltext{1}{Harvard-Smithsonian Center for Astrophysics, 60 
Garden Street, Cambridge, MA 02138 USA; asozzett@cfa.harvard.edu}
\altaffiltext{2}{INAF - Osservatorio Astronomico di Torino, 10025 Pino
Torinese, Italy}
\altaffiltext{3}{Department of Physics, and Kavli Institute for Astrophysics 
and Space Research, Massachusetts Institute of Technology, Cambridge, MA 02139 USA}
\altaffiltext{4}{Department of Physics \& Astronomy,
Bowling Green State University, Bowling Green, OH 43403 USA}
\altaffiltext{5}{California Institute of Technology, 1200 East California 
Boulevard, Pasadena, CA 91125, USA}
\altaffiltext{6}{NASA Postdoctoral Program Fellow, Goddard Space Flight Center, 
8800 Greenbelt Rd Code 690.3, Greenbelt MD 20771 USA}
\altaffiltext{7}{Lowell Observatory, Flagstaff, AZ, USA}
\altaffiltext{8}{Planetary Science Institute, Tucson, AZ 85719 USA}
\altaffiltext{9}{Instituto de Astrof\'\i sica de Canarias, 38200 La Laguna, Tenerife, Spain}
\altaffiltext{10}{Las Cumbres Observatory Global Telescope, Goleta, CA 93117}
\altaffiltext{11}{Department of Physics, University of California, Santa Barbara, CA 93106}
\altaffiltext{12}{Alfred P.\ Sloan Research Fellow}



\begin{abstract}

We report new spectroscopic and photometric observations of the parent
stars of the recently discovered transiting planets \mbox{TrES-3} and
\mbox{TrES-4}.  A detailed abundance analysis based on high-resolution
spectra yields [Fe/H] $= -0.19\pm 0.08$, $T_\mathrm{eff} = 5650\pm
75$~K, and $\log g = 4.4\pm 0.1$ for \mbox{TrES-3}, and [Fe/H] $=
+0.14\pm 0.09$, $T_\mathrm{eff} = 6200\pm 75$~K, and $\log g = 4.0\pm
0.1$ for \mbox{TrES-4}. The accuracy of the effective temperatures is
supported by a number of independent consistency checks.  The
spectroscopic orbital solution for \mbox{TrES-3} is improved with our
new radial-velocity measurements of that system, as are the
light-curve parameters for both systems based on newly acquired
photometry for \mbox{TrES-3} and a reanalysis of existing photometry
for \mbox{TrES-4}. We have redetermined the stellar parameters taking
advantage of the strong constraint provided by the light curves in the
form of the normalized separation $a/R_\star$ (related to the stellar
density) in conjunction with our new temperatures and
metallicities. The masses and radii we derive are $M_\star =
0.928_{-0.048}^{+0.028}~M_{\sun}$, $R_\star =
0.829_{-0.022}^{+0.015}~R_{\sun}$, and $M_\star =
1.404_{-0.134}^{+0.066}~M_{\sun}$, $R_\star =
1.846_{-0.087}^{+0.096}~R_{\sun}$ for \mbox{TrES-3} and
\mbox{TrES-4}, respectively.
With these revised stellar parameters we obtain improved values for
the planetary masses and radii. We find $M_p =
1.910_{-0.080}^{+0.075}~M_\mathrm{Jup}$, $R_p =
1.336_{-0.036}^{+0.031}~R_\mathrm{Jup}$ for \mbox{TrES-3}, and $M_p =
0.925 \pm 0.082~M_\mathrm{Jup}$, $R_p =
1.783_{-0.086}^{+0.093}~R_\mathrm{Jup}$ for \mbox{TrES-4}. We
confirm \mbox{TrES-4} as the planet with the largest radius among the
currently known transiting hot Jupiters.

\end{abstract}





\section{Introduction}
\label{sec:introduction}

The 40 transiting planet systems confirmed as of August 
2008\footnote{For a complete listing see {\tt
http://www.inscience.ch/transits/}, or {\tt http://exoplanet.eu}~.}
show a remarkable diversity of properties, which is indicative of the
complexity of planet formation and evolution processes. Many different
follow-up studies enabled by the special orientation of these systems
\citep[e.g.,][see Charbonneau et al.\ 2007 for a review]{queloz00,
charbon02, charbonneau05, knutson07, tinetti07} have brought about
rapid improvements in evolutionary models of planet interiors and
atmospheres \citep{baraffe04, lecavelier06, guillot06, burrows07}.
The increasing predictive power of these models is beginning to drive
even more challenging observations. On the other hand, the precision
and accuracy with which the most basic planet properties such as the
mass and radius can be determined is currently limited by our
knowledge of the properties of the host stars.  Significant
uncertainties remain in the stellar mass and radius determinations of
many systems. In some cases this is due to poorly determined
photospheric properties (mainly temperature and metallicity), and in
others to a lack of an accurate luminosity estimate. Additionally, the
variety of methodologies used for these determinations and the
different approaches towards systematic errors have resulted in a
rather inhomogeneous set of planet properties, as discussed by
\cite{torres08}. This complicates the interpretation of patterns and
correlations that are being proposed \citep[e.g.,][]{Mazeh:05,
guillot06, Hansen:07}. Recent improvements in the analysis techniques
have the potential to increase the accuracy of the stellar and
planetary parameters significantly, especially for the large majority
without a direct distance estimate. In particular, the application of
the constraint on the stellar density that comes directly from the
transit light curves has been shown to be superior to the use of other
indicators of luminosity such as the surface gravity ($\log g$)
determined spectroscopically \citep{sozzetti07, holman07}.
\cite{torres08} have recently reanalyzed a large subset of the known
transiting planets, incorporating these improvements and applying a
uniform methodology to all systems.

In the present work we focus on two of the recently discovered
transiting systems, \mbox{TrES-3} \citep{ODonovan:07} and
\mbox{TrES-4} \citep{Mandushev:07}, which lack accurate estimates for
the photospheric properties of the parent stars and as a result have
more uncertain stellar and planetary parameters.  To improve upon
these properties, we present new radial-velocity and photometric
observations of \mbox{TrES-3} with which we refine both the light
curve solution and the spectroscopic orbit. We also carry out a
reanalysis of existing \mbox{TrES-4} photometry utilizing a technique 
which treats stellar limb darkening using adjustable parameters. 
We perform the first detailed spectroscopic determination of the photospheric
properties of both stars, and we make use of the constraint on the
stellar density mentioned above to infer more accurate values for the
stellar and planetary masses and radii. Our paper is organized as
follows.  In \S\,\ref{sec:obs} we summarize the observations.  In
\S\,\ref{sec:atmpar} we present our effective temperature and
metallicity determinations, along with several consistency checks
aimed at establishing the accuracy of the temperatures.  In
\S\,\ref{sec:orb} we report an updated spectroscopic orbital solution
for \mbox{TrES-3}, and the light curve solutions for both systems are
discussed in \S\,\ref{lcanalysis}.  Section\,\ref{sec:physics} then
describes our determination of the stellar masses and radii, which in
turn lead to refined values for the planetary parameters over those
reported in the discovery papers. We conclude in \S\,\ref{sec:summ} by
providing a summary of our results and by discussing whether the
properties of the host stars give any useful clues on the origin of
the strongly contrasting densities of their close-in gas giant
planets, particularly in comparison with the other known transiting
planet systems.

\section{Observations}
\label{sec:obs}

\subsection{New radial velocities for \mbox{TrES-3}}
\label{sec:rvs}

High-resolution, high-SNR (signal-to-noise ratio) spectroscopic
observations of \mbox{TrES-3} were obtained in July and September 2007
with the HIRES spectrograph on the Keck~I telescope \citep{vogt94},
using essentially the same setup as in the discovery paper. The
spectra cover the effective wavelength range $\sim$3200--8800\,\AA,
and an iodine (I$_2$) gas absorption cell placed in front of the
spectrograph slit was used to superimpose a dense set of narrow
molecular lines between $\sim$5000\,\AA\ and $\sim$6000\,\AA. This
iodine spectrum provides a stable wavelength reference and a means of
monitoring the instrumental profile that are crucial for achieving
precise radial velocity determinations.  The observations were reduced
and extracted using the MAKEE software written by T.\ Barlow.  We
obtained 4 spectra with typical exposure times of 15 min resulting in
SNRs of $\sim$100 pixel$^{-1}$, which we combined with the discovery
observations reported by \cite{ODonovan:07} for a total of 11 iodine
spectra.

Precise radial velocities were measured following a procedure based on
the methodology developed for the AFOE spectrograph~\citep{korz00},
and adapted for the processing of HIRES spectra.  Conceptually, the
method models the observed star and I$_2$ spectrum by using templates
for the stellar and I$_2$ spectra. This model includes physically
motivated parameters that describe the spatial and temporal variations
of the instrument, including the instrumental profile, as well as the
sought-after relative Doppler shift with respect to the stellar
template. The model parameters are adjusted to minimize the difference
between the model and the observations in a least-squares sense, down
to the Poisson noise.  The stellar template is estimated by
deconvolving an observation taken without superimposing I$_2$, whereas
the template for the I$_2$ is based on a high-resolution and high SNR
scan of the gas cell using the Fourier Transform Spectrometer on the
McMath Solar Telescope at the Kitt Peak
Observatory~\citep[see][]{butler96}.  Each spectral order is modeled
as a whole but analyzed independently.  The final radial velocities
are estimated from the mean of the Doppler shifts computed for each
order and their uncertainties from the standard deviation of that
mean. The resulting velocities, expressed in the solar system
barycentric frame, along with their associated formal errors are
reported in Table~\ref{rvtres3}.  These measurements include and
supersede the velocities presented in the discovery paper, and they
also correct a minor error in the previously published dates of observation.  
The revised orbital solution is presented and discussed in \S\,\ref{sec:orb}.

\subsection{Differential photometry: new \mbox{TrES-3} data and revisited \mbox{TrES-4} data}
\label{sec:newphot}

In addition to the photometric measurements presented in the discovery
paper \citep{ODonovan:07}, we have collected six other high-cadence,
high-precision transit light-curves of \mbox{TrES-3}. One was obtained
in the $V$ band on UT 2007 April 23 using the CCD camera of the IAC80
telescope at the Observatorio del Teide, Tenerife, Spain, and the
other five were gathered using KeplerCam \citep[see,
e.g.,][]{Holman:06} at the FLWO 1.2-m telescope: two in the Sloan $g$
and $r$ bands on UT 2007 April 25 and 2008 March 27, and the remaining
three in the $i$ band on UT 2008 March 9, April 12, and May 8.  On all
nights conditions were quite good, except for the presence of high
cirrus and highly variable seeing on UT 2008 April 12. The transits
were observed at airmasses ranging between $\sim$1.00 and $\sim$1.80.

All datasets were reduced using standard calibration techniques (overscan 
correction, trimming, bias subtraction, flat fielding). We then 
performed aperture photometry of \mbox{TrES-3} and between 10 and 30 comparison 
stars, depending on the filter and exposure time. We experimented with different 
choices for the aperture size, comparison star ensemble, and weighting of 
the comparison stars, aiming for the smallest out-of-transit (OOT) RMS.
In practice, the best aperture size was approximately twice the FWHM
of the stellar image on each night, and the best results were obtained
using a straight average of the normalized light curves of the
comparison stars. As for the formal errors on each photometric data
point, we used the product of the OOT RMS of each light-curve and the
factor $\beta\gtrsim 1$, used to account for departures from Gaussian
(``white'') uncorrelated noise \citep[see, e.g.,][]{winn08b}.

For consistency, we also re-processed in the same way the two
discovery light-curves of \mbox{TrES-3} presented in O'Donovan et
al. (2007), as well as the three photometric datasets utilized by
Mandushev et al. (2007) in their discovery announcement of
\mbox{TrES-4}.

\begin{figure} 
\centering
\includegraphics[width=1.\textwidth]{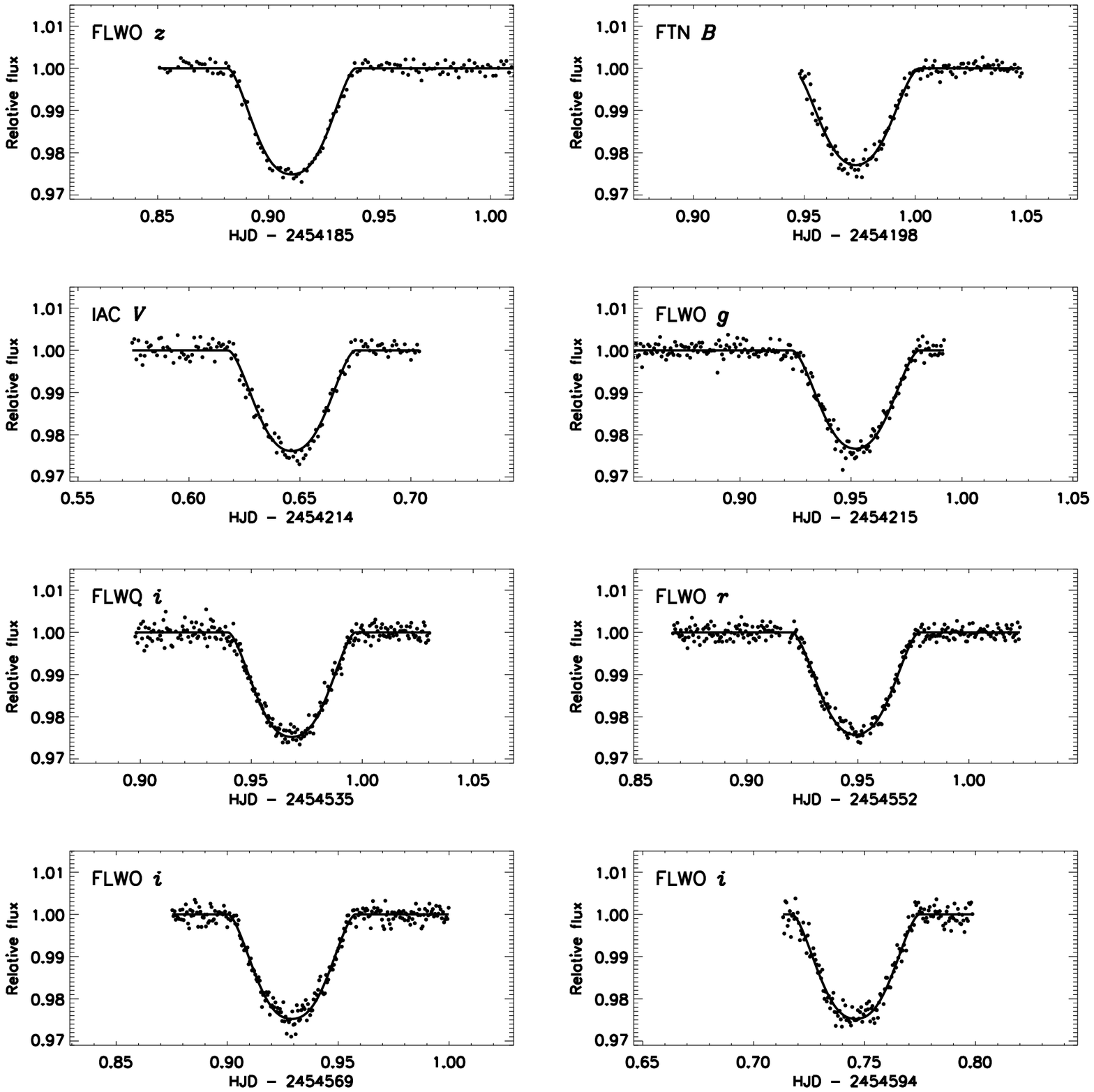} 
\caption{Relative flux of the \mbox{TrES-3} system as a function of time from the center 
of transit, adopting the ephemeris in Table~\ref{tab:par_tres3}. 
Each of the light-curves is labeled with the telescope and filter employed. 
We have overplotted the simultaneous best-fit solution, adopting the appropriate 
quadratic limb-darkening parameters for each band pass (see text for details).\label{tres3phot}} 
\end{figure}

\begin{figure} 
\centering
\includegraphics[width=1.\textwidth]{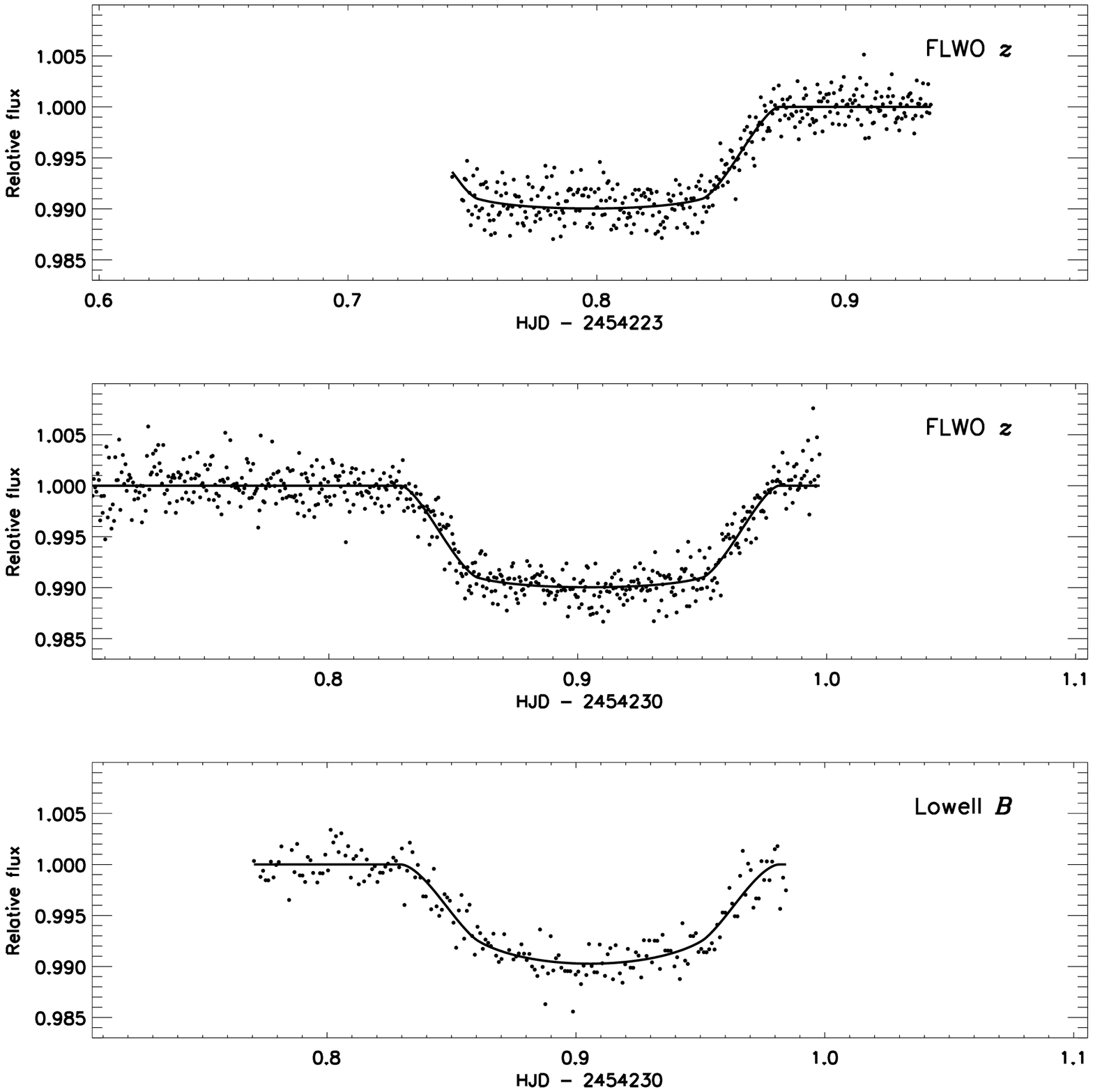} 
\caption{Relative flux of the \mbox{TrES-4} system as a function of time from the center 
of transit, adopting the ephemeris in Table~\ref{tab:par_tres4}. 
Each of the light-curves is labeled with the telescope and filter employed. 
As in Figure~\ref{tres3phot}, we have overplotted the simultaneous best-fit model, 
adopting the appropriate quadratic limb-darkening parameters for each band pass (see text 
for details).\label{tres4phot}} 
\end{figure}

The main characteristics of the \mbox{TrES-3} and \mbox{TrES-4}
light-curves are summarized in Table~\ref{tbl:lctres3}.  The final set
of photometric time-series of \mbox{TrES-3} in all filters (including
the discovery data) is available in machine-readable form in the
electronic version of Table~\ref{tbl:phottres3}, and is plotted in
Figure~\ref{tres3phot}. Table~\ref{tbl:phottres4} reports all
photometric data for \mbox{TrES-4}, which are shown graphically in
Figure~\ref{tres4phot}.

\section{Atmospheric parameters and age constraints} 
\label{sec:atmpar}

\subsection{Spectroscopic abundance analysis} 
\label{sec:keck}

A detailed abundance analysis was carried out using the Keck/HIRES
template spectra of \mbox{TrES-3} and \mbox{TrES-4}. The stellar
atmosphere parameters $T_\mathrm{eff}$, $\log g$, and [Fe/H] were
determined using standard methodology \citep[e.g.,][]{gonzlamb96,
gonzalez01, santos04}, which we summarize as follows. Equivalent
widths (EWs) for a set of relatively weak \ion{Fe}{1} and \ion{Fe}{2}
lines were measured manually in the Keck spectra using the {\tt
splot\/} task in IRAF~\citep[see, e.g.,][and references
therein, for details on the specific choice of lines]{sozzetti04}.
These were then used together with a grid of Kurucz ATLAS
plane-parallel stellar model atmospheres \citep{kurucz93} as inputs to
the 2002 version of the MOOG spectral synthesis code
\citep{sneden73}\footnote{\tt
http://verdi.as.utexas.edu/moog.html\,.}.  Atmospheric parameters were
derived under the assumption of local thermodynamic equilibrium (LTE),
imposing excitation and ionization balance. Formal uncertainties on
$T_\mathrm{eff}$, $\log g$, and microturbulent velocity $\xi_t$ were
derived using the approach described in \cite{neuforge97} and
\cite{gonzvant98}, while the nominal uncertainty for [Fe/H]
corresponds to the scatter obtained from the \ion{Fe}{1} lines rather
than the formal error of the mean.  The resulting set of parameters
for \mbox{TrES-3} is $T_{\rm eff} = 5650 \pm 75$~K, $\log g = 4.4 \pm
0.1$, $\xi_t = 0.85 \pm 0.05$ \kms, and [Fe/H] = $-0.19 \pm 0.08$, and
for \mbox{TrES-4} we obtain $T_{\rm eff} = 6200 \pm 75$~K, $\log g =
4.0 \pm 0.1$, $\xi_t = 1.50 \pm 0.05$ \kms, and [Fe/H] = $+0.14 \pm
0.09$.  These values are collected in Table~\ref{star_tres3} and
Table~\ref{star_tres4}, respectively.

\begin{figure}
\centering
$\begin{array}{c}
\includegraphics[width=.38\textwidth]{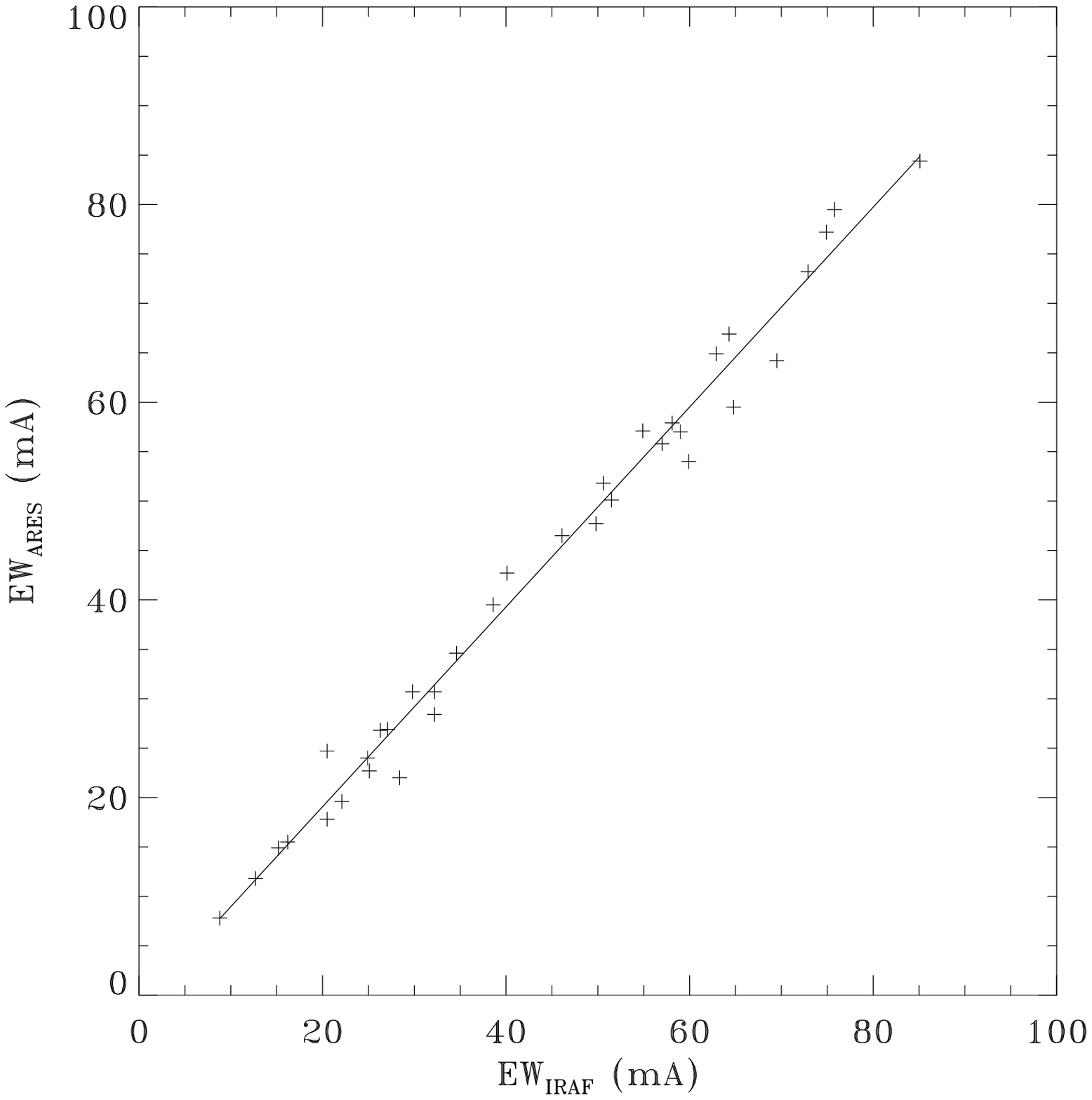} \\
\includegraphics[width=.38\textwidth]{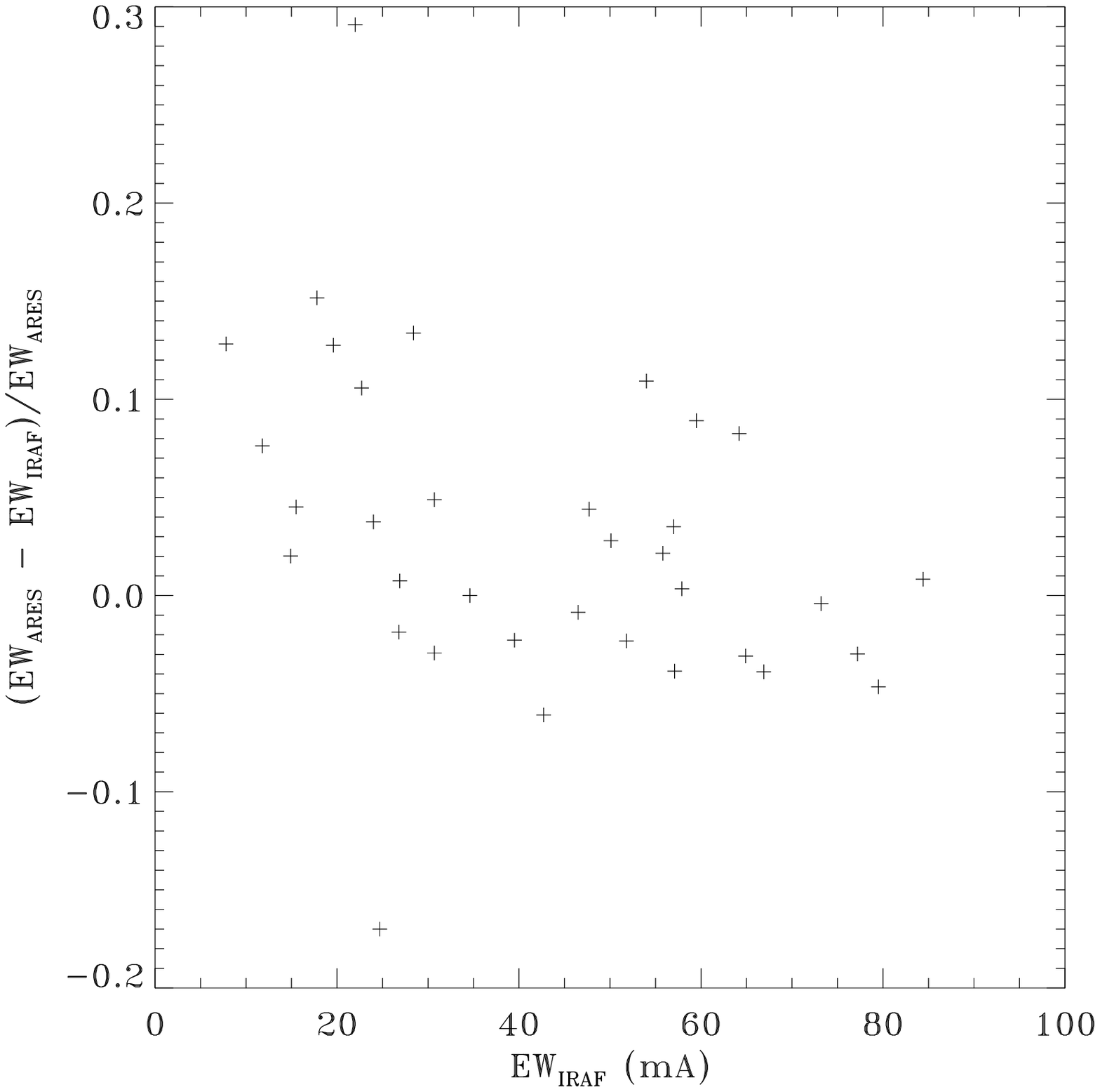} \\
\includegraphics[width=.38\textwidth]{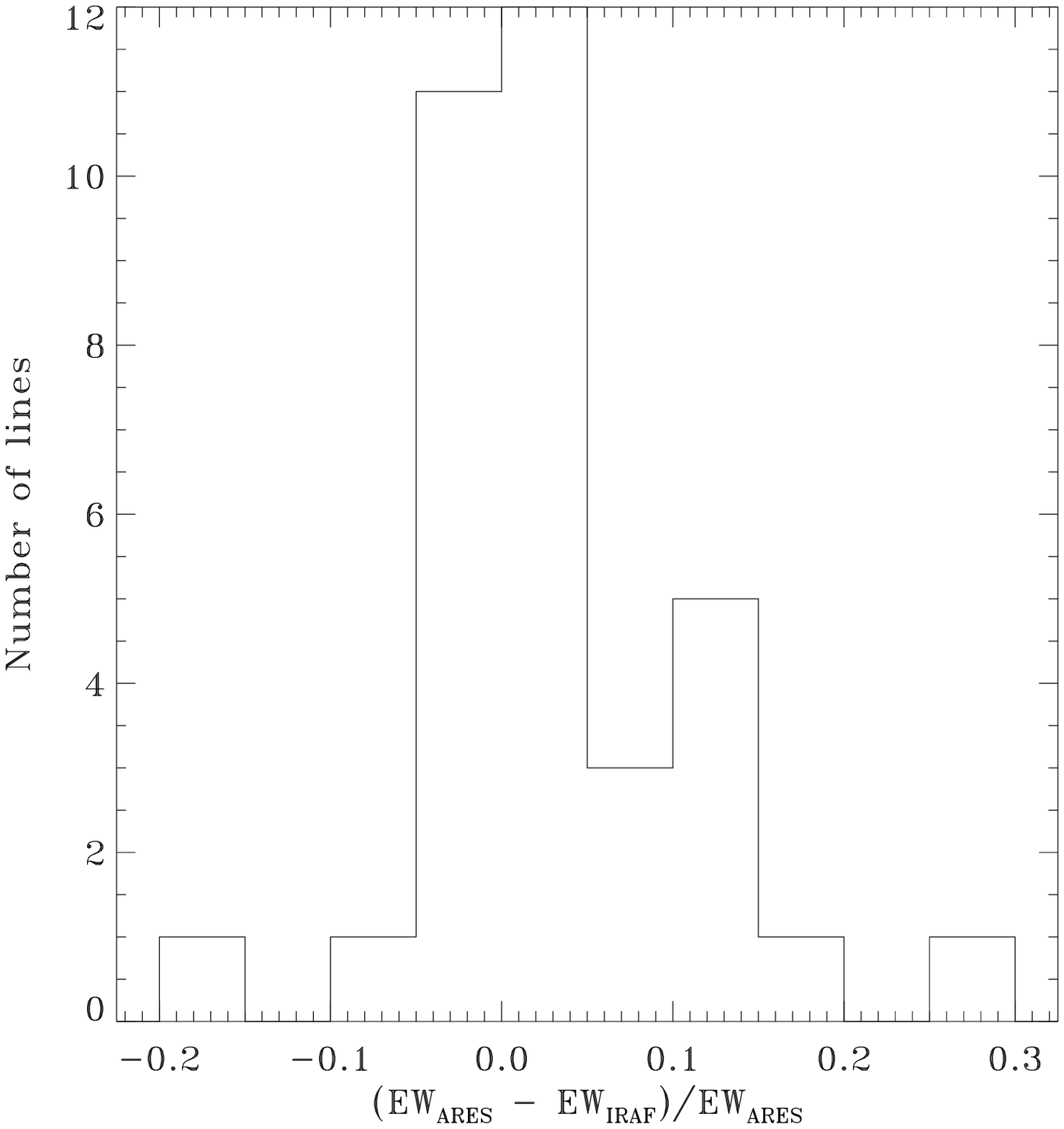} \\
\end{array}$
\caption{Top: EW of selected \ion{Fe}{1} lines in the \mbox{TrES-3} template spectrum 
measured manually with IRAF vs. EWs measured automatically with ARES. 
Center: the fractional difference between the two measurements as a
function of EW. Bottom: Histogram of the fractional differences. \label{aresiraf}}
\end{figure}

As a check, the manual EW measurements used above were compared
against the results from the automated software ARES\footnote{\tt
http://www.astro.up.pt/$^\sim$sousasag/ares\,.}, made available to the
community by \cite{sousa07}.  Figure~\ref{aresiraf} shows the
comparison between the EWs measured for \mbox{TrES-3} using ARES and
those from the manual approach with IRAF. The top panel indicates
excellent agreement between the two; the linear fit has a slope of
1.011. There are no significant correlations with the \ion{Fe}{1} line
strength (see middle panel): the fractional difference between EWs
measured by the two methods has a probability of no correlation of
0.1. Finally, the histogram in the bottom panel shows there is no
appreciable systematic difference between the methods. The mean
difference and scatter are only 0.67 m\AA\, and 2.6 m\AA,\,
respectively, which are smaller than found by \cite{sousa07} from a
similar comparison between FEROS, HARPS and UVES spectra.
Consequently, the results of the abundance analysis with MOOG using
EWs measured with ARES are virtually identical to those presented
earlier.

Following \cite{gonzalez08} we also synthesized a number of unblended
\ion{Fe}{1} lines in the template spectra of both \mbox{TrES-3} and
\mbox{TrES-4}, and determined projected rotational velocities of
$v\sin i = 1.5 \pm 1.0$ \kms\ and $v\sin i = 8.5 \pm 0.5$ \kms,
respectively.

Our new determinations of the atmospheric parameters for the parent
stars of \mbox{TrES-3} and \mbox{TrES-4} are generally in good
agreement with the values reported by \cite{ODonovan:07} and
\cite{Mandushev:07}, the main difference being that those authors
assumed [Fe/H] = 0.0 in their studies, whereas we find significant
departures from solar metallicity in both stars.

\subsection{External checks on $T_{\rm eff}$}
\label{sec:checks}

We describe here additional estimates of the effective temperature for
\mbox{TrES-3} and \mbox{TrES-4} that serve to test the accuracy of our
determinations above.

\subsubsection{CfA spectroscopy}
\label{sec:cfa}

Both planet host stars were observed spectroscopically with the Center
for Astrophysics (CfA) Digital Speedometer \citep{Latham:92} as part
of the regular follow-up after the discovery of photometric signals
suggesting transits. These observations cover 45~\AA\ in a single
echelle order centered at 5187~\AA, and have $\lambda/\Delta\lambda
\approx 35,\!000$.  While this resolving power is moderately high, the
SNRs of the spectra are low, ranging from 7 to 13 per resolution
element of 8.5~\kms\ for the 13 exposures of \mbox{TrES-3}, and SNRs
of 11--13 for the 7 exposures of \mbox{TrES-4}. Nevertheless, useful information
on the stellar properties can be extracted from these spectra as
described by \cite{Torres:02}. Briefly, the observed spectra are
cross-correlated against a library of synthetic spectra based on
Kurucz model atmospheres \citep[see][]{Nordstrom:94, Latham:02},
calculated over a wide range of values of $T_{\rm eff}$, $\log g$,
[Fe/H], and rotational velocity $V_\mathrm{rot}$. The combination of
parameters yielding the highest correlation averaged over all
exposures is adopted as a representation of the properties of the
star. Due to the narrow wavelength range and limited SNRs, the first
three of the above properties are typically strongly correlated and
are difficult to determine simultaneously. Given these constraints, 
in the discovery papers we initially held the metallicity fixed at 
the solar value for both stars.

For \mbox{TrES-3}, which our Keck spectroscopy indicates is slightly
metal-poor, we repeated the above determination using a fixed
metallicity of [Fe/H] $= -0.5$, and interpolated $T_{\rm eff}$ and
$\log g$ to the precise composition from Table~\ref{star_tres3}. We
obtained $T_{\rm eff} = 5530 \pm 130$~K and $\log g = 4.45 \pm 0.17$,
where the uncertainties include contributions both from the error in
the adopted [Fe/H] and the internal errors. The temperature as well as
the surface gravity are consistent with the Keck determinations,
within the uncertainties. Similarly for \mbox{TrES-4} we repeated the
determinations for [Fe/H] $= +0.5$, given the metal-rich composition
indicated by the Keck spectroscopy, and interpolated to the
intermediate value from Table~\ref{star_tres4}.  The results are
$T_{\rm eff} = 6270 \pm 150$~K and $\log g = 3.96 \pm 0.17$, which are
once again in good agreement with the more reliable estimate from
Keck.

\subsubsection{Line depth ratios}
\label{sec:ldr}

While the temperature determinations discussed so far are implicitly
based on the \emph{strength} of the spectral lines, \cite{Gray:91}
have demonstrated that highly precise information can also be
extracted using the \emph{ratio} of the depths of two spectral lines
having different sensitivity to temperature. However, rather than
yielding absolute temperatures, this technique in its original
formulation only measures changes in temperature, albeit with
extremely high precision often reaching a few Kelvin \citep[see,
e.g.,][]{Gray:94, Catalano:02, Kovtyukh:03}.  Absolute temperatures
can still be obtained with recourse to external color-temperature
calibrations, since line-depth ratios (LDRs) are usually strongly
correlated with the color index of the star. The accuracy of such
$T_{\rm eff}$ determinations is then limited by the calibrations
themselves.

\cite{Biazzo:07} have presented LDR-$T_{\rm eff}$ calibrations based
on 26 carefully selected lines of Fe, V, Sc, Si, and Ni in the
spectral interval 6190--6280~\AA, grouped into 16 line pairs. These
calibrations are valid for stars with temperatures between
$\sim$3800~K and $\sim$6000~K, so unfortunately the technique is not
applicable to \mbox{TrES-4}. For \mbox{TrES-3} we measured all 26
lines in our high-SNR Keck template spectrum, and adopted the
\cite{Biazzo:07} calibrations appropriate for stars with rotational
velocities of 10~\kms\ (very close to the $v\sin i$ value reported by
\cite{Mandushev:07} and obtained also in this work).  The temperature
scale of these relations relies on a transformation between $B\!-\!V$
and $T_{\rm eff}$ by \cite{Gray:05}, which makes use of a mixture of
dwarf and giant temperatures obtained by many different methods and
does not account for differences in metallicity. For the present work
we have preferred to use more sophisticated color-temperature
relations such as those by \cite{Ramirez:05} and \cite{Casagrande:06},
which account not only for luminosity class but include also
metallicity terms, and are based on effective temperatures derived
homogeneously by the Infrared Flux Method.  The conversion from the
LDR-based $T_{\rm eff}$ inferred from the \cite{Biazzo:07} relations
back to an average color for the star using the \cite{Gray:05}
prescription gives $B\!-\!V = 0.641 \pm 0.007$. Our two preferred
color-temperature relations mentioned above then yield a weighted
average temperature of $T_{\rm eff} = 5710 \pm 70$~K for
\mbox{TrES-3}, in which the uncertainty includes observational errors
propagated from the LDR measurements as well as the scatter of the
calibrations. This result is consistent with our more direct estimate
in \S\,\ref{sec:keck}.

\subsubsection{H$_\alpha$ line profiles}
\label{sec:halpha}

\begin{figure}
\plottwo{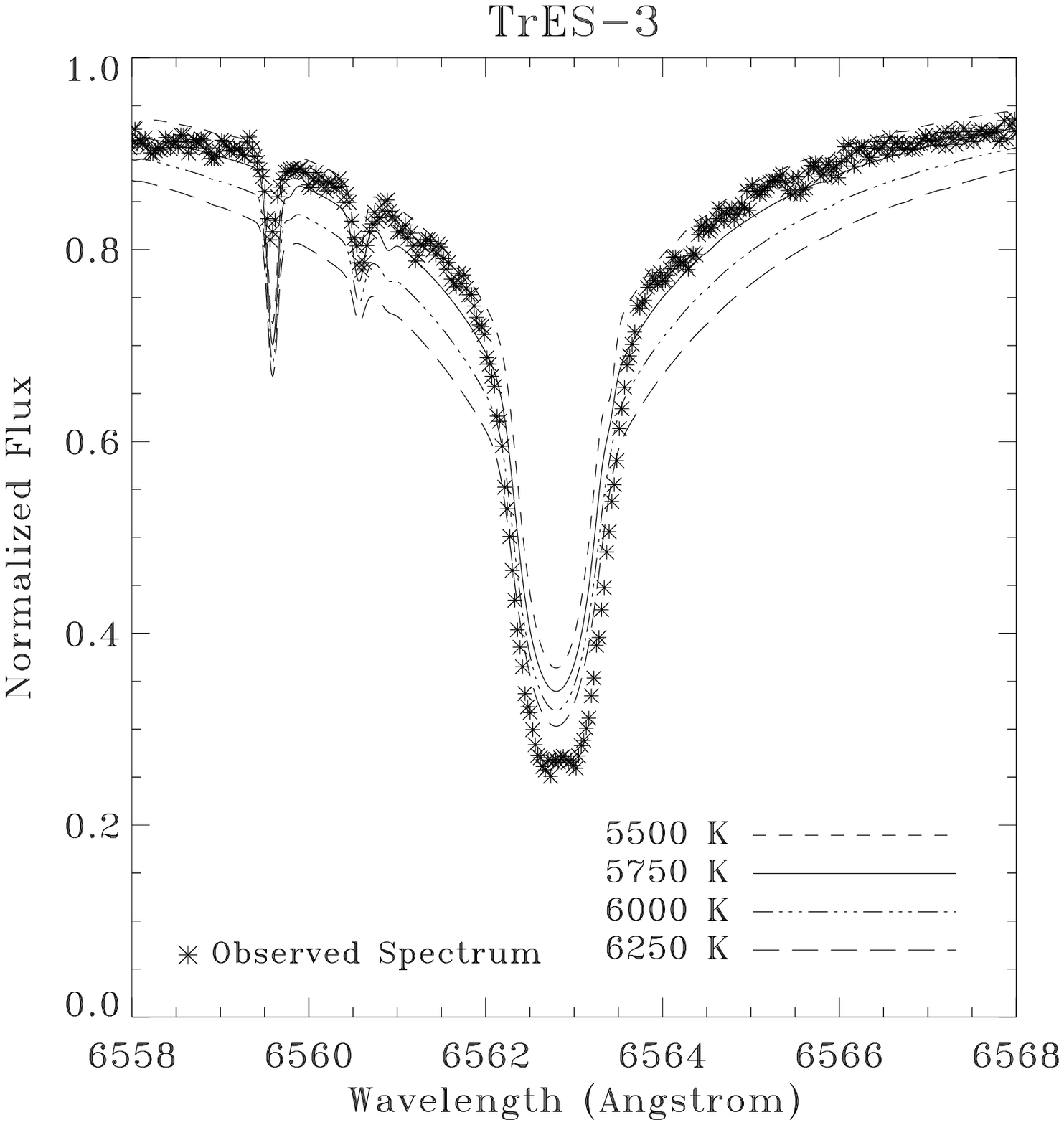}{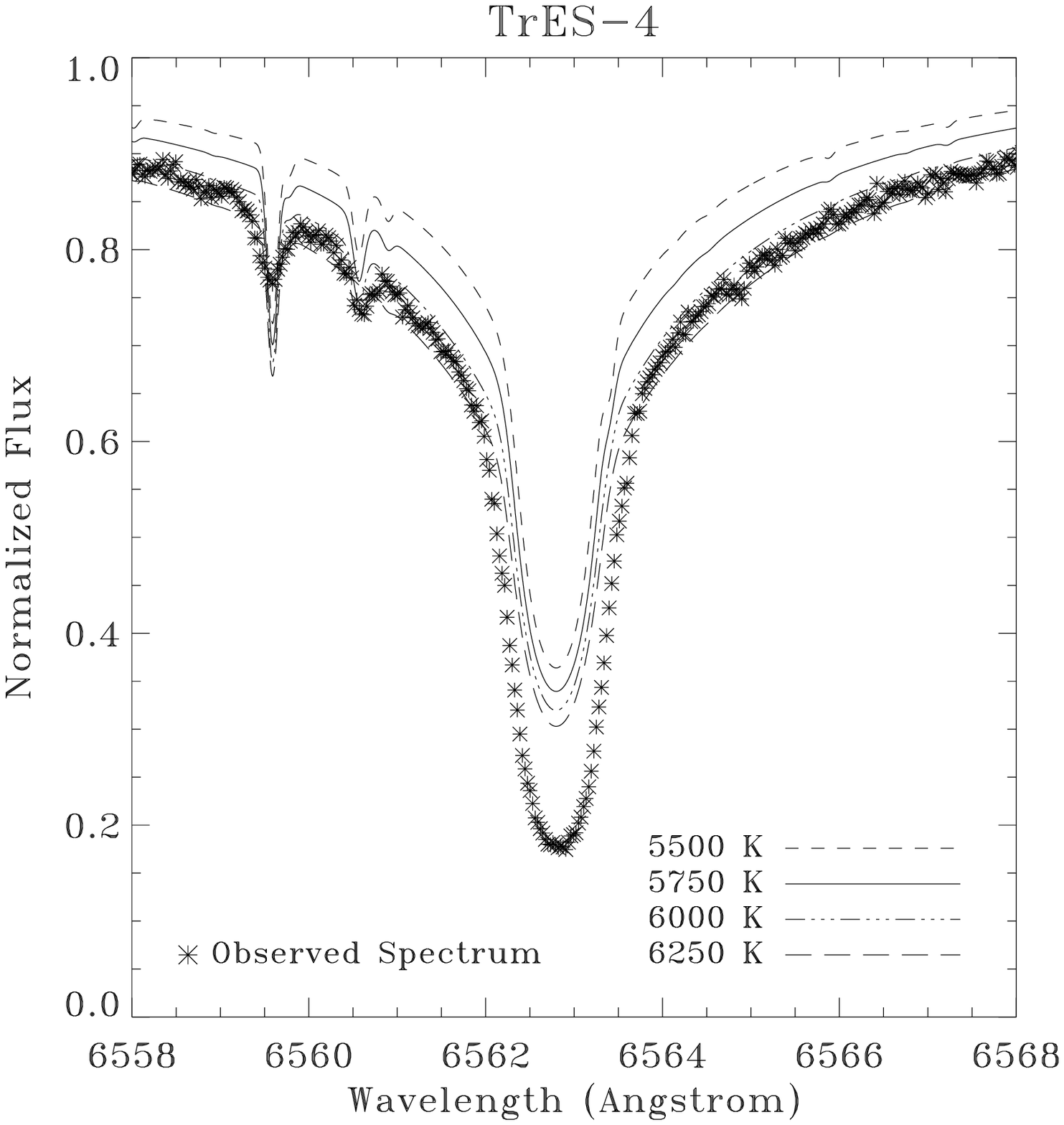} 
\caption{Observed H$_\alpha$ profile in the Keck template spectrum of \mbox{TrES-3} 
(left) and \mbox{TrES-4} (right) compared with four synthetic spectra with 
[m/H] = 0.0, $\log g= 4.5$, and effective temperatures of 5500, 5750, 6000, and 6250 K, 
respectively. \label{halpha}} \end{figure}

As is well known, the wings of the $H_\alpha$ line (but not its core,
formed higher up in the atmosphere under non-LTE conditions) are very
sensitive to $T_\mathrm{eff}$ variations, but are relatively
insensitive to changes in $\log g$ and [Fe/H] \citep[see, e.g.,][and
references therein]{sozzetti07, santos06}. This allows for a useful
consistency check on our $T_\mathrm{eff}$ estimates above. We compared
the observed H$_\alpha$ line profiles in our Keck template spectra of
\mbox{TrES-3} and \mbox{TrES-4} against synthetic profiles for
solar-metallicity dwarfs ([Fe/H] = 0.0, $\log g = 4.5$) from the
Kurucz database. The results of this exercise are displayed in
Figure~\ref{halpha}, in which 10~\AA\ regions centered on H$_\alpha$
are shown for each star together with four calculated profiles for
different values of $T_\mathrm{eff}$. In both cases the temperatures
one would infer from these comparisons agree well with the estimates
reported in Table~\ref{star_tres3} and Table~\ref{star_tres4}.

\subsubsection{Photometric estimates}
\label{sec:phot_teff}

An additional check on the effective temperatures is available from
the multi-color photometry for \mbox{TrES-3} and
\mbox{TrES-4}. Measurements in Johnson $BV$, Cousins $RI$, and 2MASS
$JHK_s$ were used to derive seven color indices for \mbox{TrES-3}, and
nine for \mbox{TrES-4} when considering also the $B_{\rm T}$ and
$V_{\rm T}$ measurements from the Tycho-2 catalog \citep{Hog:00}. The
calibrations by \cite{Ramirez:05} and \cite{Casagrande:06} then
yielded weighted average temperatures of $5390 \pm 50$~K and $6135 \pm
50$~K for \mbox{TrES-3} and \mbox{TrES-4}, respectively, ignoring
extinction and adopting [Fe/H] in each case as determined in
\S\,\ref{sec:keck}. The uncertainties include photometric errors and
metallicity errors, as well as the scatter of the calibrations, but
exclude unquantified systematic errors in the calibrations themselves.
These temperatures are 260~K and 65~K cooler than our spectroscopic
determinations above. They can be reconciled with the values in
\S\,\ref{sec:keck} if we assume the presence of reddening, and correct
each of the indices. For \mbox{TrES-4} the required value of
$E(B\!-\!V)$ is hardly significant ($0.013 \pm 0.010$ mag), but for
\mbox{TrES-3} we obtain $E(B\!-\!V) = 0.071 \pm 0.013$
mag.\footnote{The presence of reddening in \mbox{TrES-3} is already
apparent from our results in \S\,\ref{sec:ldr}, in which the LDR-based
$B\!-\!V$ color is significantly bluer than the measured value of
$B\!-\!V = 0.712 \pm 0.009$ \citep{ODonovan:07}.}  For comparison, the
reddening maps of \cite{Burstein:82} and \cite{Schlegel:98} indicate a
\emph{total} reddening along the line of sight to each star of only
$E(B\!-\!V) \sim 0.028$ mag, and similar results are obtained from the
model of Galactic dust distribution by~\cite{drimmel01}. This
is not inconsistent with the small value we infer for \mbox{TrES-4},
but it is much smaller than our estimate for \mbox{TrES-3}.  Possible
explanations include patchy interstellar material combined with the
relatively coarse resolution of these maps \citep[a few arc minutes
for][]{Schlegel:98}, or perhaps the presence of circumstellar material
in \mbox{TrES-3}.

\subsection{Constraints on the age}
\label{sec:age}

The reliability of age indicators for stars older than 1--2~Gyr, such
as chromospheric activity and lithium (Li) abundance, as well as their
interagreement, have been the subject of much debate in the literature
\citep[e.g.,][]{pace04, lambred04, song04, sestito05, sestito06}. The
difficulties are due to many factors, including (but not limited to)
non-trivial correlations between chromospheric activity, rotation,
mass, and age, limited availability of activity estimates averaged
over entire stellar activity cycles\footnote{One can appreciate how
sensitive the age can be to activity cycles by considering the
temporal evolution of the activity levels in the Sun, with values of
\logrhk\ that range from $-5.10$ to $-4.75$. These correspond to ages
of $\sim 8$~Gyr and $\sim$ 2.5~Gyr, respectively \citep[see,
e.g.,][]{Henry:96}.}, insufficient understanding of the temporal
evolution of Li depletion due to the complex interplay between various
processes (e.g., convection, mixing, diffusion, mass loss), and
non-negligible differences in the observed behavior of chromospheric
activity and Li depletion as a function of mass, age, and chemical
abundance between stars in young and old clusters, and in the field.

We have nonetheless attempted to use the \ion{Ca}{2} activity
indicator and the lithium abundance as measured in our HIRES spectra
of \mbox{TrES-3} and \mbox{TrES-4} to provide independent constraints
for comparison with the formal age estimates determined below from
evolutionary models (see \S\,\ref{sec:physics}), as well as to search
for possible peculiarities of these two planet hosts compared to other
samples of stars with and without planets.

\begin{figure}
\plotone{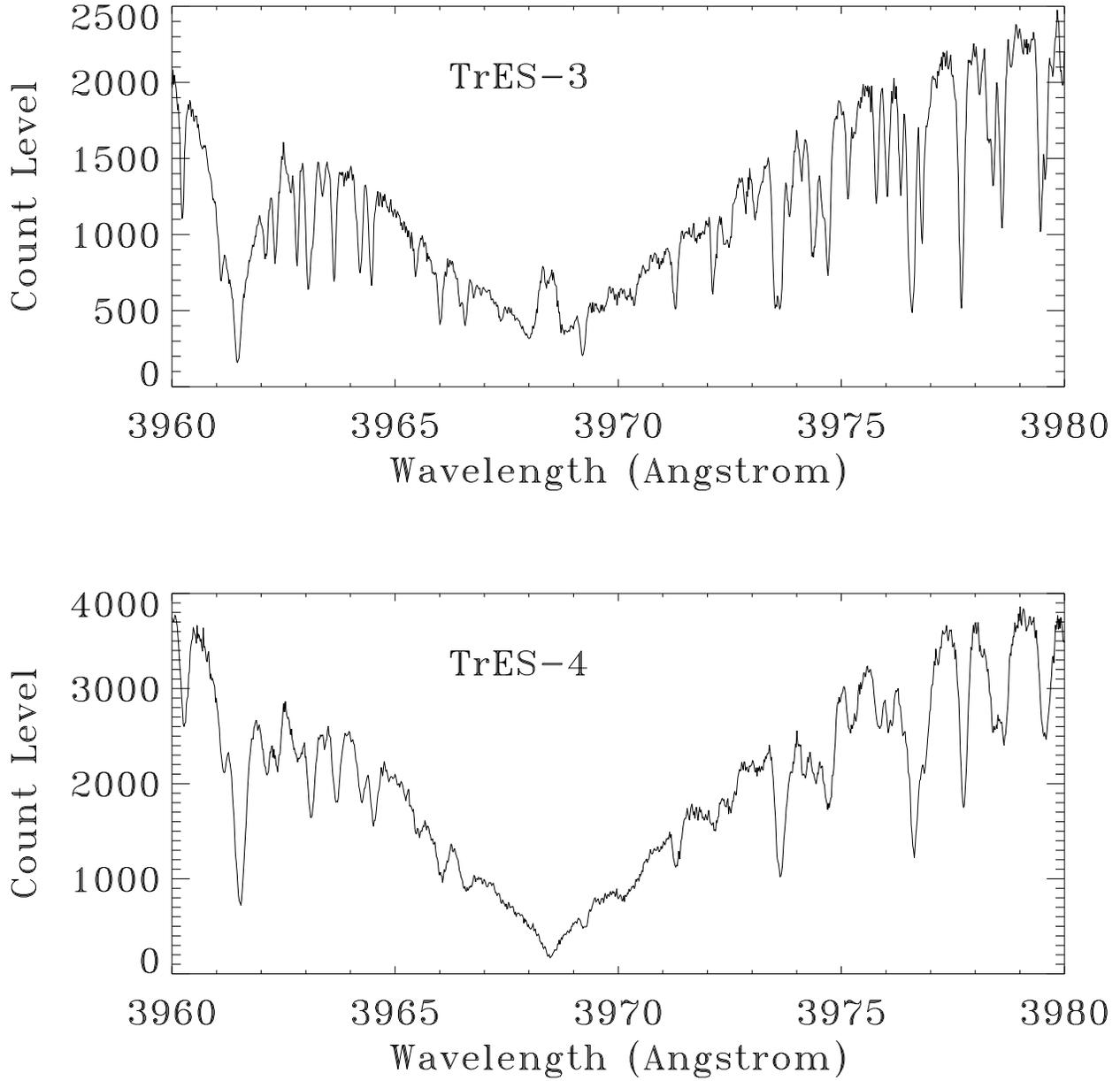}
\caption{Top: A 20 \AA\ region of the Keck template spectrum of \mbox{TrES-3}
centered on the \ion{Ca}{2} H line.  Bottom: The same, but for \mbox{TrES-4}. 
\label{ca_tres3_4}}
\end{figure}

Figure~\ref{ca_tres3_4} shows a region of the HIRES template spectra
for each star centered on the \ion{Ca}{2} H line.  Clear emission is
seen in the core of the line of \mbox{TrES-3}, but not in
\mbox{TrES-4}. Following the procedure outlined in \cite{sozzetti04}
we measured the chromospheric emission ratio $\log R^\prime_{HK}$,
corrected for the photospheric contribution, from the \ion{Ca}{2} H
and K lines in each of our spectra. We obtained $\langle\log
R^\prime_{HK}\rangle = -4.54\pm0.13$ and $\langle\log
R^\prime_{HK}\rangle = -5.11\pm0.15$ for \mbox{TrES-3} and
\mbox{TrES-4}, respectively. \mbox{TrES-3} thus appears moderately
active, while \mbox{TrES-4} is quite the opposite. The resulting
chromospheric age estimates, using the relations summarized in
\cite{wright04}, are $t = 0.9 \pm 0.7$ Gyr for \mbox{TrES-3} and $t =
9.4 \pm 1.7$ Gyr for \mbox{TrES-4}.

\begin{figure}
\plotone{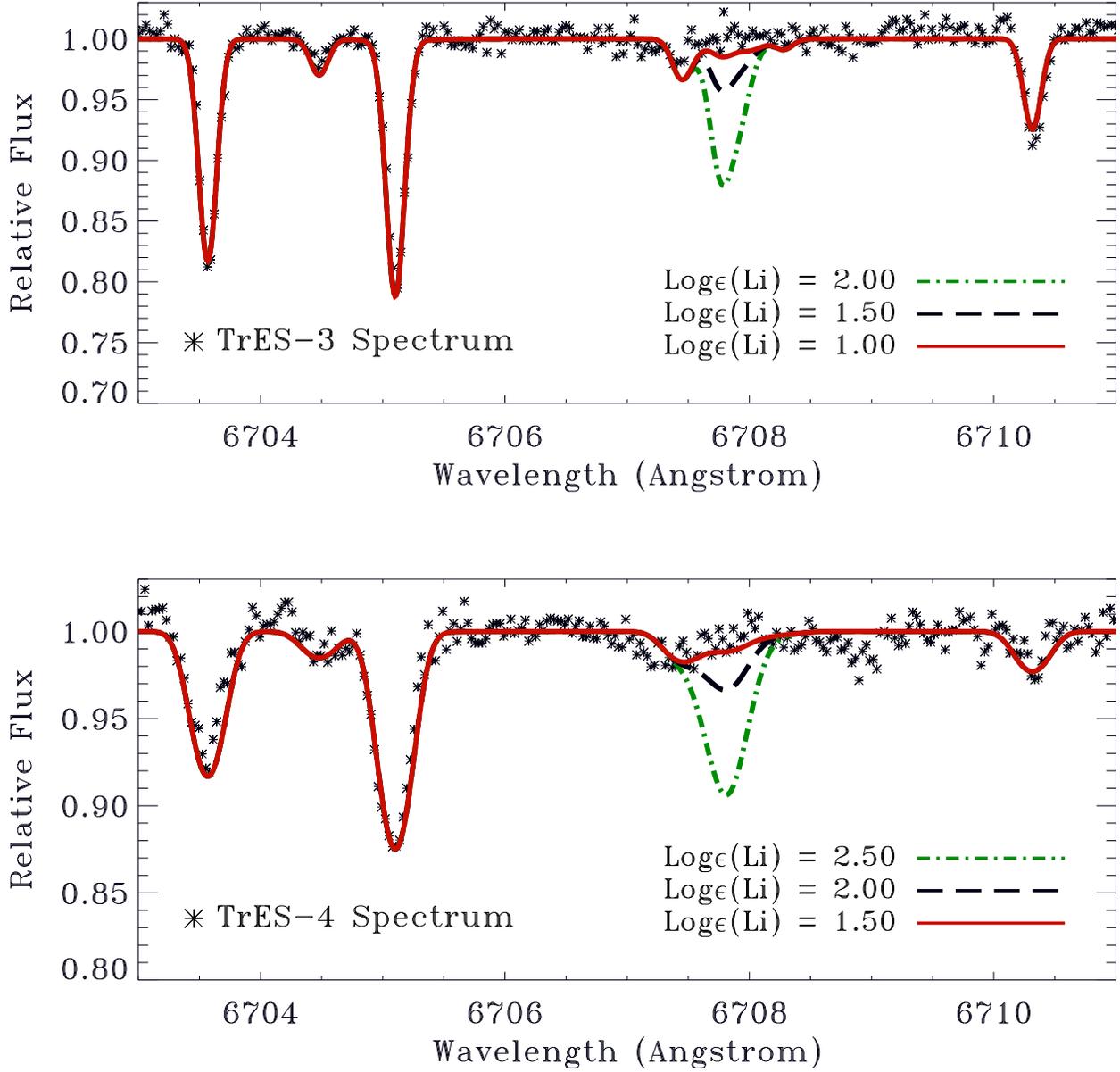}
\caption{Top: A 10 \AA\ region of the Keck template spectrum of \mbox{TrES-3} 
containing the \ion{Li}{1} line at 6707.8 \AA\ (filled dots), 
compared to three synthetic profiles (lines of various colors and styles), 
each differing only in the lithium abundance assumed. Bottom: The same, 
but for \mbox{TrES-4}.\label{licomp}}
\end{figure}

In Figure~\ref{licomp} we show the results of the spectral synthesis
of a 10~\AA\ region centered on the Li~$\lambda$6707.8 line, using the
atmospheric parameters derived from the Fe-line analysis and the line
list of \cite{reddy02}. The two panels display the comparison between
the observed spectra and three synthetic spectra, each differing only
in the assumed Li abundance. Neither star shows a measurable Li line,
and we can only place upper limits of $\log\epsilon{\rm (Li)} < 1.0$
and $\log\epsilon{\rm (Li)} < 1.5$ for \mbox{TrES-3} and
\mbox{TrES-4}, respectively. By comparison with average Li abundance
curves as a function of effective temperature for clusters of
different ages \citep{sestito05}, one would infer a rather old age for
\mbox{TrES-3} of $t\gtrsim 4$ Gyr, nominally inconsistent with the
estimate above from the \ion{Ca}{2} activity index. Based on its Li
the object appears decidedly older than Hyades stars of the same
$T_\mathrm{eff}$, and this argument is corroborated
\citep[e.g.,][]{pace04} by the small $v\sin i$ we measure (see
Table~\ref{star_tres3}). For \mbox{TrES-4}, the age inferred from the
\cite{sestito05} relations is $t > 5$ Gyr, which is in qualitative
agreement with the absence of significant chromospheric activity as
well as with the inferred value of the surface gravity (see
Table~\ref{star_tres4}), which indicates the star is somewhat evolved.
The measured Li abundance for \mbox{TrES-3} is in line with values
determined by \cite{israelian04} for a sub-sample of nearby planet
hosts with the same $T_\mathrm{eff}$, while \mbox{TrES-4} appears more
depleted than other planet hosts of similar $T_\mathrm{eff}$, $\log
g$, and [Fe/H]. We discuss briefly in \S\,\ref{sec:summ} some
implications of these findings in the broader context of the existence
of chemical peculiarities in planet hosts in comparison with stars
without detected planets.

From the measurements in this section we conclude that the above
results are consistent with the notion that neither star is very young
($t >$ 1--2 Gyr), that \mbox{TrES-3} is likely to be of intermediate
age ($\sim$3~Gyr), and that \mbox{TrES-4} appears older and more
evolved.

\section{Revised spectroscopic orbital solution for \mbox{TrES-3}}\label{sec:orb}

\begin{figure}
\plotone{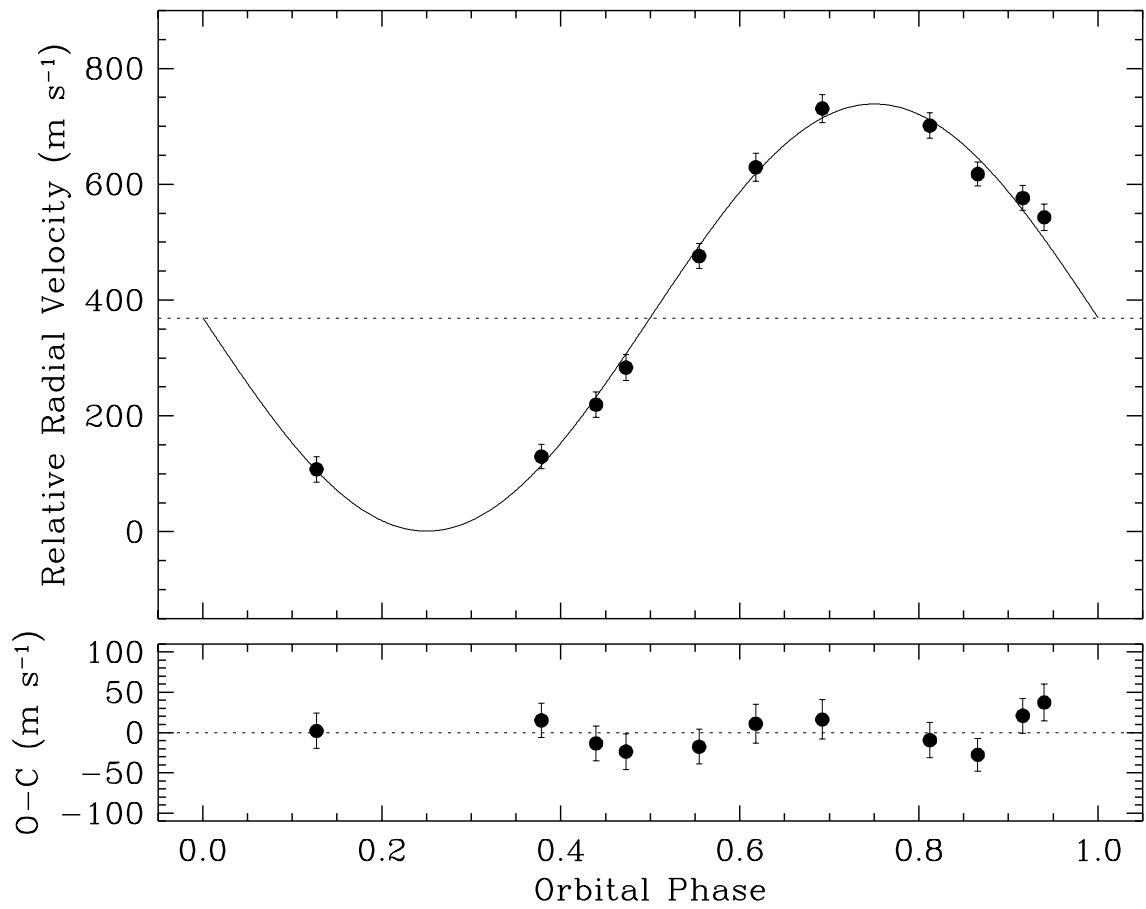}
\vskip -0.5in
\caption{Revised spectroscopic orbital solution for
TrES-3, with the post-fit residuals shown at the bottom. \label{specorb_tres3_4}}
\end{figure}

Using the radial velocities presented in \S\,\ref{sec:rvs} we have
updated the orbital solution given by \cite{ODonovan:07}, adopting the
improved ephemeris described below in \S\,\ref{ttv}.  A Keplerian
circular orbit was adjusted to the data, and the results may be seen
in Table~\ref{tab:par_tres3} and are shown graphically in
Figure~\ref{specorb_tres3_4}. As found also by \cite{ODonovan:07},
the solution gives a scatter that is larger than expected from the
internal velocity errors.  This is most likely due to velocity
``jitter'' associated with chromospheric activity.  If we model this
as excess scatter to be added quadratically to the internal errors, we
find that the amount of jitter required to produce a reduced $\chi^2$
value near unity is $\sim$18.5~\ms.  External estimates of the jitter
for \mbox{TrES-3} can be made on the basis of the spectral type, the
measured value of $v \sin i = 1.5 \pm 1.0$~\kms, and the activity
index $\log R^{\prime}_{\rm HK} = -4.54 \pm 0.13$, and vary
considerably but generally range from about 5 to 20~\ms\
\citep{Saar:98, Santos:00, Paulson:02, Wright:05}.  These estimates
are thus consistent with our findings.  Nevertheless, as a test we
also modeled the data with an eccentric orbit but did not obtain much
improvement, and the result for the eccentricity was not significantly
different from zero ($e = 0.015 \pm 0.019$).  There are no indications
of long-term variations in the observations at hand.

For \mbox{TrES-4} we adopt in the following the spectroscopic orbit by
\cite{Mandushev:07}, since the radial-velocity material has not
changed. For clarity, the orbital solution is reported again in Table~\ref{tab:par_tres4}.

\section{Light-curve analysis} \label{lcanalysis}

The analysis of the differential photometry for \mbox{TrES-3} and
\mbox{TrES-4} was carried out in essentially the same manner as
described in \cite{torres08}. We refer the reader to that paper for
the details.  Given that we have multiple light curves for each
system, we first determined the individual times of transit as
described below in order to improve the ephemeris, and we subsequently
refined the light-curve parameters.

\subsection{Transit timings}\label{ttv}

\begin{figure} 
\centering
\includegraphics[width=1.\textwidth]{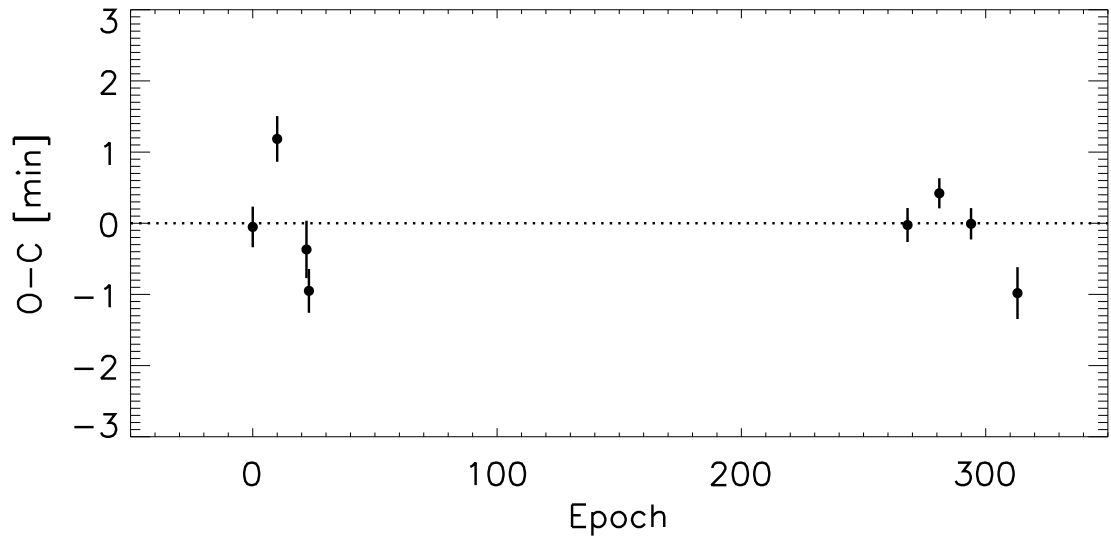} 
\caption{Timing residuals (observed - calculated) for eight observed transits of TrES-3,
according to the ephemeris derived in this work. \label{trestiming}} 
\end{figure}

To determine the center of each measured transit event we initially
adopted the light-curve parameters from the discovery papers.  For
\mbox{TrES-3}, we adopted a quadratic limb darkening (LD) law with
coefficients from \cite{Claret:04}, interpolated to the values
$T_\mathrm{eff} = 5650$ K, $\log g = 4.40$ dex, [M/H] $= -0.2$ dex,
and $\xi_t = 2.0$ \kms. All LD coefficients for the various filters are
reported in Table~\ref{tbl:ldcoeff}.  We fitted each light curve
individually to solve only for the time of transit center $T_c$ and
also a linear function of time (two parameters) to describe the OOT flux
(the slope in the OOT flux accounts for systematic errors, including
differential extinction). The measured mid-transit times are presented
in Table~\ref{midtrans}.  We then fitted a straight line to the
central times of the eight transits of the form $T_c(E) = T_c(0) + E\cdot P$
and derived the following new ephemeris (which we report in
Table~\ref{tab:par_tres3}): $T_c=2454185.9104\pm0.0001$ (HJD),
$P=1.30618581\pm0.00000051$ days.  With a number of degrees of freedom
$\nu=6$, the resulting reduced $\chi^2/\nu=5.87$ indicates a rather
poor fit.  The transit timing residuals for \mbox{TrES-3} are shown in
Figure~\ref{trestiming}. There are four outliers at the 2--3.5$\sigma$
level. This could be seen as evidence suggesting that the period is
not constant. Alternatively, the errors might have been
underestimated.  The data available are not enough to draw any
significant conclusion on the nature of these variations, but clearly
additional observations are warranted. We leave for a future study the
evaluation of the dynamical interpretation and significance of the
transit times, as was done, for example, by \cite{steffen05} for
TrES-1 and by \cite{diaz08} for OGLE-TR-111b.

Similarly for \mbox{TrES-4}, we re-fitted the three discovery
light-curves using the \cite{Claret:04} LD coefficients listed in
Table~\ref{tbl:ldcoeff} appropriate for $T_\mathrm{eff} = 6200$ K,
$\log g=4.00$ dex, [M/H]$=+0.1$ dex, and $\xi_t$ = 2.0 \kms. The
corresponding times of transit center, derived with the same procedure
adopted for \mbox{TrES-3}, are presented in
Table~\ref{midtrans}. Given the short time baseline of these
observations, we did not derive a new ephemeris for the system, but
for the purpose of the analysis presented in the next Section we
simply adopted the Mandushev et al. (2007) values of $P$ and $T_c$.

\subsection{Light-curve system parameters}\label{lcfits}

Next, for both \mbox{TrES-3} and \mbox{TrES-4} we locked the transit
times and OOT baseline functions at the values indicated above and
then, under the assumption of a circular orbit, we fitted all light
curves simultaneously using the algorithm by \cite{mandel02} to derive
the relevant quantities radius ratio $R_p/R_\star$ (where $R_p$ and
$R_\star$ are the planetary and stellar radius, respectively), 
inclination $i$, projected separation $a/R_\star$ (where $a$ is
the semi-major axis), and impact parameter $b\equiv a\cos i/R_\star$. 
For \mbox{TrES-3} we found $R_p/R_\star =
0.1654 \pm 0.0018$, $i = 81\fdg83 \pm 0\fdg12$, $a/R_\star = 5.922 \pm
0.051$, and $b = 0.840 \pm 0.010$.  For \mbox{TrES-4}, we obtained
$R_p/R_\star = 0.09964 \pm 0.00086$, $i = 82\fdg59 \pm 0\fdg40$,
$a/R_\star = 5.93 \pm 0.19$, and $b = 0.766 \pm 0.020$.  These
estimates were derived assuming initial values for the stellar and
planetary masses of $M_\star = 0.936$ $M_\odot$ and $M_p = 1.920$
$M_{\rm Jup}$ for \mbox{TrES-3}, and $M_\star = 1.394$ $M_\odot$ and
$M_p = 0.923$ $M_{\rm Jup}$ for \mbox{TrES-4}, although the results
are insensitive to these values.

Recent studies of transiting exoplanets light curves \citep[e.g.,][and
references therein]{south08} have highlighted some issues related to
how much effect different treatments of limb darkening can have on the
light-curve solutions.  In particular, these do not appear to be
significantly affected by the specific choice of the LD law, while
fixing the LD coefficients to their theoretical values seems to result
in significantly smaller uncertainties in the fitted parameters with
respect to the case in which the LD coefficients are also adjusted
during the fitting procedure. The above results were obtained keeping
all LD coefficients fixed. We decided to investigate the effect of
fitting for the LD coefficients in the following way. Given the
quality of the data available to us, we can only fit for one
coefficient (or one combination of the two coefficients). We chose to
fix the quadratic coefficient $u_2$ and solve for the linear
coefficient $u_1$. We found that allowing complete freedom in the
linear coefficient resulted in ``unphysical'' solutions (e.g.,
largely negative coefficients). \cite{south08} found that the
fitted LD coefficients are usually within 0.1--0.2 of the theoretical
Claret values, and that is consistent with our experience \citep[see,
e.g.,][]{winn07}.  We then decided to use an a priori constraint
enforcing an agreement of $\sim0.2$, modifying the merit function as
follows:
\begin{equation}
\chi^2 = \sum_j\left[\frac{f_j^{obs}-f_j^{calc}}{\sigma_j}\right]^2+\left(\frac{u-u_1}{0.2}\right)^2\, ,
\end{equation}
where $f_j^{obs}$ is the stellar flux observed at time $j$, $\sigma_j$
its corresponding error, $f_j^{calc}$ is the model value, $u$ is the
adjustable linear LD coefficient, and $u_1$ its theoretical value
(appropriate for each band-pass). Furthermore, for \mbox{TrES-3} all
three $i$-band light curves were required to agree on the LD
parameter, and the same requirement was set on the two $z$-band light
curves for \mbox{TrES-4}.  The results for the \mbox{TrES-3} system
parameters in this case were $R_p/R_\star = 0.1655 \pm 0.0020$, $i =
81\fdg85 \pm 0\fdg16$, and $a/R_\star = 5.926 \pm 0.056$. For
\mbox{TrES-4}, we obtained $R_p/R_\star = 0.09921 \pm 0.00085$, $i =
82\fdg59 \pm 0\fdg40$, and $a/R_\star = 5.94 \pm 0.21$.

In both cases, the agreement between the system parameters derived
keeping the LD coefficients fixed and those when the LD coefficients
are part of the solution was excellent, with only a slight increase in
the estimated uncertainties for the latter case. Indeed, as already
noted by \cite{south08}, fixing the LD coefficients at their
theoretically predicted values does not appear to significantly bias
the results. However, in the interest of providing more conservative error estimates, 
we believe that, for the purpose of the analysis of
high-quality light-curves such as the ones presented in this paper, 
a procedure that treats LD coefficients as adjustable parameters is preferable. 
Based on the above considerations, for both \mbox{TrES-3} and \mbox{TrES-4} we elected to include in
Table~\ref{tab:par_tres3} and Table~\ref{tab:par_tres4} the values of
the systems parameters obtained from the light-curve analysis in the
case in which the linear LD coefficient was allowed to float. We note
that this approach provides a departure from the \cite{torres08}
analysis.

\section{Stellar and planetary parameters}\label{sec:physics}

The revised spectroscopic determinations of $T_{\rm eff}$ and [Fe/H],
along with the new spectroscopic orbital solution for \mbox{TrES-3}
and the light curve fits presented above, allow us to refine the
determination of the stellar and planetary properties for both
systems. To establish the properties of the parent stars we rely on
stellar evolution models by \cite{Yi:01} and \cite{Demarque:04}. We
make explicit use of the constraint on the stellar density provided by
the light-curve quantity $a/R_{\star}$, as described by
\cite{sozzetti07}. The procedure follows closely that given in the
previous citation.  Briefly, we seek the best match (in a $\chi^2$
sense) between the measured \{$T_{\rm eff}$, [Fe/H], $a/R_{\star}$\}
and points on a finely interpolated grid of isochrones spanning a wide
range of metallicities and ages. Two minor improvements in this
procedure, described in more detail by \cite{torres08}, have to do
with the weighting of each point sampled along the isochrones
according to the distance in $T_{\rm eff}$-[Fe/H]-$a/R_{\star}$ space
compared to the observed values, and an additional weighting according
to the likelihood that the star is in a particular evolutionary
state. The latter effect is accounted for by multiplying the first
weight by the expected number density of stars at each location in the
H-R diagram, according to an adopted Initial Mass Function (which in
this case is simply a power law with a Salpeter index). The impact of
these weighting factors is generally minor.

The results for the stellar properties of \mbox{TrES-3} and
\mbox{TrES-4} are presented in Table\,\ref{star_tres3} and
Table\,\ref{star_tres4}. The distance estimate to \mbox{TrES-3}
accounts for interstellar extinction ($A_V = 3.1 \cdot E(B\!-\!V) =
0.22 \pm 0.04$ mag), as described in \S\,\ref{sec:phot_teff}.  For
both stars the surface gravity inferred from the models is typically
much more accurate than the spectroscopic gravity determination,
showing the power of the constraint on the luminosity and size of the
star afforded by $a/R_{\star}$. The results for both \mbox{TrES-3} and
\mbox{TrES-4} supersede those given recently by \cite{torres08}
because of the new photometry and radial velocities contributed here
in the former case, and the new light-curve solutions for both
systems. The evolutionary ages for the two stars are qualitatively
consistent with our conclusions from \S\,\ref{sec:age}, but are
somewhat more uncertain in the case of \mbox{TrES-3}. Finally, using
the distance estimates inferred in this work, the UCAC2 proper motion
components~\citep{zacharias04} reported by~\cite{ODonovan:07}
and~\cite{Mandushev:07}, and the mean radial velocity values
$RV_\mathrm{TrES-3}=+9.58\pm0.73$~\kms\, and
$RV_\mathrm{TrES-4}=-16.40\pm0.19$~\kms\, as measured with the CfA
Digital Speedometers, we obtain Galactic space motion vectors [$U$,
$V$, $W$] = [+27.3, +6.7, +33.0]~\kms\, and [$U$, $V$, $W$] =
[$-$43.9, $-$39.1, $-$6.9]~\kms\, for \mbox{TrES-3} and
\mbox{TrES-4}, respectively (where $U$ is taken to be positive toward the Galactic
anticenter). We collect these results along with the other properties
derived previously in Table~\ref{star_tres3} and Table~\ref{star_tres4}.

The planet parameters follow from the stellar properties and the
results of the transit light curve and spectroscopic orbits, and are
presented in Table\,\ref{tab:par_tres3} and
Table\,\ref{tab:par_tres4}. For \mbox{TrES-3}, our planetary mass and
radius are significantly larger compared with the determinations in
the discovery paper and in \cite{torres08}. This is due in part to the
increased mass and radius for the parent star, but also to the larger
radius ratio based on the new photometry.

\section{Summary and discussion}\label{sec:summ}

Our detailed spectroscopic analyses of \mbox{TrES-3} and \mbox{TrES-4}
have yielded accurate values of the atmospheric properties
($T_\mathrm{eff}$ and [Fe/H]), which are critical for establishing the
fundamental properties of the hosts. The accuracy of the temperatures
is supported by a number of independent checks (low-resolution
spectroscopy, line-depth ratios, H$_\alpha$ line profiles,
color-temperature calibrations) that gives us confidence that the
inferred stellar properties are reliable. We find that \mbox{TrES-3}
is a main-sequence G dwarf with a metallicity about 1.5 times lower
than the Sun's, not a very common occurence among exoplanets hosts. 
\mbox{TrES-4} is a somewhat evolved late F star that is nearing the
end of its main-sequence phase, and is slightly enhanced in its iron
content with respect to the solar abundance.

The agreement we find between age indicators for \mbox{TrES-3} and
\mbox{TrES-4} based on measurements of the \ion{Ca}{2} activity
levels, the lithium abundances, and rotation, and the evolutionary age
inferred from the models is fair, although, as discussed in
\S\,\ref{sec:age}, the reliability of empirical age estimates for
stars that are not young ($t\gtrsim 1$ Gyr) is somewhat questionable.
The model estimates themselves are not without their problems.
Nevertheless, we conclude the stars are 1--3 Gyr old. Neither star
stands out as peculiar when compared with other planet hosts with
similar physical properties. In \mbox{TrES-3} the small $v\sin i$
value and the fact that we can only place an upper limit on the Li
abundance are consistent with the notion that planet hosts with
$T_\mathrm{eff}$ similar to the Sun appear to rotate more slowly and
are more Li-depleted than stars without detected planets. As pointed
out recently \citep[][and references therein]{gonzalez08}, this
evidence suggests that a planet-forming disk may induce additional
rotational braking, leading to enhanced mixing in the stellar
envelope, which in turn accelerates the destruction of
lithium. \mbox{TrES-4}, on the other hand, does not seem consistent
with the claim by \cite{gonzalez08} that hotter planet hosts with
$T_\mathrm{eff} \gtrsim 6100$~K have higher Li abundances, possibly
due to self-enrichment processes. It is worth keeping in mind,
however, that these discrepancies may not be significant given the
large spread in the Li abundance for field stars with the temperature
and mass of \mbox{TrES-4} \citep[e.g.,][]{lambred04}. New
investigations on these issues are clearly needed based on uniform
analyses of large samples of planet hosts and statistically
significant, well-defined control samples of stars without detected
planets.

New radial-velocity measurements for \mbox{TrES-3} presented here have
enabled us to revise the spectroscopic orbit for that system. We
detect no indication of any longer-term variations in the radial
velocities that might suggest the presence of another body in the
system.  However, the small number of observations and their limited
time span of only 6 months emphasize the need for continued Doppler
monitoring of this and other transiting planet systems to investigate
the possibility of additional companions.  In stark contrast to the
considerable ground-based and space-based efforts invested in studying
in great detail the atmospheric properties of many of these objects,
which have undoubtedly led to tremendous insights into their
structure, formation, and evolution, the amount of radial-velocity
data available for transiting planets is meager, and often does not go
beyond the handful of observations published in the discovery papers.
Interest in the radial velocities seems to be quickly lost. It should
be pointed out that the frequency of close-in giant planets ($P < 10$
day) with additional massive planets in outer orbits \citep[up to the
detection limit of today's Doppler surveys, $\sim 4$ AU; see,
e.g.,][and references therein]{butler06} is about 12\% (8 out of 70
systems discovered via RV methods), and it would therefore be wise to
continue the velocity monitoring of some of the transiting systems.
If additional planets in a transiting system were detected, they might
also be found to undergo transits, and such a discovery would allow us
to constrain structural models for gas giants akin to Jupiter or
Saturn, and open exciting opportunities for additional investigations
with present ({\em Spitzer}) and upcoming (JWST) space-borne
observatories.

Accurate stellar properties for \mbox{TrES-3} and \mbox{TrES-4} have
been derived here following the approach described in
\cite{sozzetti07} and \cite{holman07}, comparing the spectroscopically
determined $T_\mathrm{eff}$ and [Fe/H] values along with the
photometrically measured $a/R_\star$ with current stellar evolution
models. These properties have in turn allowed us to refine the
determination of the mass and radius of the planets. In particular, we
confirm that \mbox{TrES-4} is the planet with the largest radius among
the currently known transiting hot Jupiters.  

Recently, Winn et al. (2008a) derived upper limits on the albedo of 
TrES-3 based on the nondetection of occultations observed at 
optical wavelengths. Our findings are relevant to this study in 
two ways.  Firstly, they strengthen the case for a circular orbit, 
which is important because if the orbit is eccentric then it would 
be possible that Winn et al. (2008a) did not observe TrES-3 at the 
actual times of occultations and that their data place no 
constraint on the albedo.  Secondly, our revised light-curve parameters 
are relevant because the upper limit on the geometric albedo ($p_\lambda$) 
was inferred from the measured upper limit on the planet-to-star flux ratio 
($\epsilon_\lambda$) according to

\begin{equation}
p_\lambda = \epsilon_\lambda(a/R_p)^2
\end{equation}

Our revised value of $(a/R_p)$ therefore leads to revised upper 
limits on the geometric albedo of TrES-3. However, this revision 
turns out to be minor: the new value of $(a/R_p)$ is only 1.5\% 
larger than the value used by Winn et al.~(2008a). The upper limits 
on the geometric albedo become weaker by about 3\%. At 99\% confidence, 
the revised upper limits are 0.31, 0.64, and 1.10 in $i-$, $z-$, and $R-$band, 
respectively.

There is a large spread in the observed radii and densities for transiting planets of
comparable mass placed at similar orbital distance from stars of very
similar properties ($T_\mathrm{eff}$, $\log g$, [Fe/H], and age). 
For example, if we consider \mbox{TrES-4} along with 5 other transiting
planet hosts (excluding \mbox{HAT-P-2}) with similar characteristics
(HD~149026, HD~209458, OGLE-TR-56, OGLE-TR-132, and WASP-1), the
nominal masses of the attending planets vary by a factor of $\sim3.5$,
but reported densities vary by a factor of $\sim7.5$
\citep[e.g.,][]{torres08}.  Many theoretical mechanisms have been
proposed to inflate the radius of a strongly irradiated planet, such
as additional sources of internal heating due to stellar insolation
\citep{guillot02}, tidal heating due to non-zero eccentricity caused
by gravitational interaction with an outer companion
\citep{bodenheimer01} or by rotational obliquity \citep{winn05},
elevated interior opacity due to enhanced atmospheric metallicity
\citep{guillot06, burrows07}, or varying core masses
\citep{fortney07}. None of these appear capable of explaining the
observed spread in density in a natural way \citep[see
also][]{Fabrycky:07}.  

Of the 6 stars just mentioned, all are more metal-rich than
the Sun, except for HD~209458, which has [Fe/H] close to solar. Some
of the above models \citep{guillot06, burrows07} predict a positive
correlation between the inferred planetary core mass and the host
star's metallicity, as in the framework of the core-accretion model of
giant planet formation \citep[e.g.,][]{pollack96, alibert05}. This
idea assumes [Fe/H] closely tracks the metallicity of the
protoplanetary disk, so that more metal-rich stars should be orbited
by more metal-rich planets, with a larger heavy-element content.
However, among these 6 systems only one planet (HD~149026b) has an
inferred core mass significantly larger than zero, and four of the
other planets have measured radii so large that the cores are likely
to be insignificant. In fact, even in the absence of a core the
observed radii cannot be reproduced by the models, with \mbox{TrES-4}
being the extreme case. Interestingly, over 40\% of the transiting
planets reported in Table~5 of \cite{torres08} do not appear to
require any core at all to explain their sizes \citep[according to the
models of][]{fortney07}.  Taking all this into consideration, the
claimed evidence for a core mass--metallicity correlation could indeed
be seen as supporting the more widely accepted scenario of formation
by core accretion, but from the indications above it may also be that
a significant fraction of these objects formed in a different way
\citep[e.g.,][and references therein]{durisen07}. 

Given the evidence collected so far, we suggest that simply
connecting the host star's characteristics to the structural
properties of transiting planets may in fact be an
over-simplification. We conclude by stressing the importance of
refining our understanding of the complex interplay between the disk
environment and a forming giant planet, and its evolutionary history
after envelope accretion, which might turn out to be more directly
responsible for its final structure and composition than the metal
content of the parent star.

\acknowledgments

AS gratefully acknowledges the Kepler mission for partial support
under NASA Cooperative Agreement NCC 2-1390. 
GT acknowledges partial support for this work from NASA Origins grant NNG04LG89G. 
DC is supported in part by NASA Origins grant NNG05GJ29G. 
FTOD acknowledges partial support for this work provided through the NASA
Postdoctoral Program at the Goddard Space Flight Center, administered by Oak
Ridge Associated Universities through a contract with NASA. JBL gratefully
acknowledges support from NSF grant AST-0307340.  Some of the data
presented herein were obtained at the W.\ M.\ Keck Observatory, which
is operated as a scientific partnership among the California Institute
of Technology, the University of California and the National
Aeronautics and Space Administration. The Observatory was made
possible by the generous financial support of the W.\ M.\ Keck
Foundation. The authors wish to recognize and acknowledge the very
significant cultural role and reverence that the summit of Mauna Kea
has always had within the indigenous Hawaiian community. We are most
fortunate to have the opportunity to conduct observations from this
mountain. This research has made use of NASA's Astrophysics Data
System Abstract Service and of the SIMBAD database, operated at CDS,
Strasbourg, France.

\clearpage

\begin{deluxetable}{lcc}
\tablecaption{Radial velocity measurements of \mbox{TrES-3}.\label{rvtres3}}
\tablewidth{0pt} 
\tablehead{\colhead{} & \colhead{Radial Velocity} & \colhead{$\sigma_\mathrm{RV}$} \\
\colhead{~~~BJD$-$2,400,000~~~}& \colhead{(m s$^{-1}$)} & \colhead{(m s$^{-1}$)}}
\startdata
 54187.04136\dotfill &     176.0     &  \phn8.2 \\
 54187.13832\dotfill &      99.4     &  10.7 \\
 54188.02349\dotfill &     189.5     &  11.2 \\
 54188.12067\dotfill &     292.4     &  13.3 \\
 54189.01696\dotfill &  $-$331.6\phs &  \phn5.5 \\
 54189.09680\dotfill &  $-$246.4\phs &  11.4 \\
 54189.14038\dotfill &  $-$195.8\phs &  \phn7.3 \\
 54288.85394\dotfill &     241.6     &  \phn7.8 \\
 54288.98956\dotfill &     116.7     &  11.5 \\
 54289.82337\dotfill &    $-$5.9     &  10.8 \\
 54372.86102\dotfill &  $-$335.9\phs &  \phn9.5 \\
\enddata
\end{deluxetable}

\clearpage

\begin{deluxetable}{lcccccc}
\tabletypesize{\normalsize}
\tablecaption{Main characteristics of the \mbox{TrES-3} and \mbox{TrES-4} light-curves\label{tbl:lctres3}}
\tablewidth{0pt}

\tablehead{
\colhead{UT Date} & \colhead{Filter} & \colhead{Observatory} & 
\colhead{Cadence (min)} & \colhead{OOT RMS} & \colhead{Res RMS} & \colhead{$\beta$}
}

\startdata
\noalign{\vskip 0.3cm}
\multicolumn{7}{c}{TrES-3} \\
\noalign{\vskip 0.3cm}
2007-03-25 &   $z$    &      FLWO    &      1.73  &   0.0012 &  0.0011  &  1.17 \\
2007-04-08 &   $B$    &      FTN     &      1.15  &   0.0011 &  0.0014  &  1.46 \\
2007-04-24 &   $V$    &      OT      &      1.34  &   0.0016 &  0.0016  &  1.30 \\
2007-04-25 &   $g$    &      FLWO    &      0.98  &   0.0014 &  0.0015  &  1.16 \\
2008-03-09 &   $i$    &      FLWO    &      0.73  &   0.0018 &  0.0016  &  1.00 \\
2008-03-27 &   $r$    &      FLWO    &      0.73  &   0.0014 &  0.0014  &  1.07 \\
2008-04-12 &   $i$    &      FLWO    &      1.05  &   0.0015 &  0.0015  &  1.02 \\
2008-05-08 &   $i$    &      FLWO    &      0.73  &   0.0021 &  0.0020  &  1.16 \\
\noalign{\vskip 0.3cm}
\multicolumn{7}{c}{TrES-4} \\
\noalign{\vskip 0.3cm}
2007-05-03  &  $z$    &      FLWO     &     0.73 &    0.0015 &  0.0016 &   1.26 \\
2007-05-10  &  $z$    &      FLWO     &     0.73 &    0.0019 &  0.0017 &   1.23 \\
2007-05-10  &  $B$    &      Lowell   &     1.54 &    0.0015 &  0.0015 &   1.22 \\
\enddata 

\tablecomments{Column 4 gives the median spacing between exposures, in minutes. Column 5 
gives the out-of-transit root-mean-squared relative flux. Column 6 gives the residual RMS relative 
flux after subtracting the best-fitting model. Column 7 gives the scaling factor $\beta$ that 
was applied to the single-point flux uncertainties to account for red noise (see \S~\ref{sec:newphot}).}

\end{deluxetable}

\clearpage

\begin{deluxetable}{lcc}
\tabletypesize{\normalsize}
\tablecaption{Differential photometry of \mbox{TrES-3}\label{tbl:phottres3}}
\tablewidth{0pt}

\tablehead{
\colhead{HJD} & \colhead{Relative flux} & \colhead{Uncertainty}
}

\startdata
\noalign{\vskip 0.3cm}
\multicolumn{3}{c}{$z$ band (FLWO 1.2-m)} \\
\noalign{\vskip 0.3cm}
2454185.850884\dotfill   &     0.99965   &     0.00115   \\
2454185.853303\dotfill   &     0.99948   &     0.00115  \\
2454185.854495\dotfill   &     0.99852   &     0.00115   \\  
\noalign{\vskip 0.3cm}
\multicolumn{3}{c}{$B$ band (FTN 2.0-m)} \\
\noalign{\vskip 0.3cm}
2454198.948010\dotfill   &     0.99987    &    0.00109  \\
2454198.948809\dotfill   &    1.00054     &   0.00109   \\
2454198.949606\dotfill   &     0.99878    &    0.00109  \\
\noalign{\vskip 0.3cm}
\multicolumn{3}{c}{$V$ band (OT 0.8-m)} \\
\noalign{\vskip 0.3cm}
2454214.574418\dotfill  & 1.00159 & 0.00157 \\ 
2454214.575348\dotfill  & 1.00247 & 0.00157 \\
2454214.576280\dotfill  & 0.99766 & 0.00157 \\
\noalign{\vskip 0.3cm}
\multicolumn{3}{c}{$g$ band (FLWO 1.2-m)} \\
\noalign{\vskip 0.3cm}
 2454215.850393\dotfill   &     0.99951    &    0.00150  \\ 
 2454215.851076\dotfill   &     0.99912    &    0.00150  \\ 
 2454215.851771\dotfill   &     0.99842    &    0.00150 \\ 
\noalign{\vskip 0.3cm}
\multicolumn{3}{c}{$i$ band (FLWO 1.2-m)} \\
\noalign{\vskip 0.3cm}
 2454535.897759\dotfill   &     0.99906     &   0.00163  \\
 2454535.898268\dotfill   &     0.99996     &   0.00163  \\
 2454535.898766\dotfill   &     0.99755     &   0.00163   \\
\noalign{\vskip 0.3cm}
\multicolumn{3}{c}{$r$ band (FLWO 1.2-m)} \\
\noalign{\vskip 0.3cm}
 2454552.866642\dotfill   &     0.99987     &   0.00135  \\
 2454552.867140\dotfill   &     1.00181     &   0.00135 \\ 
 2454552.867638\dotfill   &     1.00086     &   0.00135  \\
\noalign{\vskip 0.3cm}
\multicolumn{3}{c}{$i$ band (FLWO 1.2-m)} \\
\noalign{\vskip 0.3cm}
 2454569.875280\dotfill   &     1.00116     &   0.00149  \\
 2454569.875778\dotfill   &     0.99975     &   0.00149  \\
 2454569.876287\dotfill   &     1.00090     &   0.00149   \\
\noalign{\vskip 0.3cm}
\multicolumn{3}{c}{$i$ band (FLWO 1.2-m)} \\
\noalign{\vskip 0.3cm}
 2454594.713269\dotfill  &      1.00015     &   0.00204  \\
 2454594.713778\dotfill  &      0.99628     &   0.00204  \\
 2454594.714276\dotfill  &      1.00394     &   0.00204   \\
\enddata 

\tablecomments{The time stamps represent the Heliocentric Julian Date
  at the time of mid-exposure. The data have been corrected for
  residual extinction effects, and the uncertainties have been
  rescaled as described in \S~\ref{sec:obs}. We intend for this table to appear in its
  entirety in the electronic version of the journal. A portion is
  shown here to illustrate its format. The data are also available
  from the authors upon request.}

\end{deluxetable}

\clearpage

\begin{deluxetable}{lcc}
\tabletypesize{\normalsize}
\tablecaption{Differential photometry of \mbox{TrES-4}\label{tbl:phottres4}}
\tablewidth{0pt}

\tablehead{
\colhead{HJD} & \colhead{Relative flux} & \colhead{Uncertainty}
}

\startdata
\noalign{\vskip 0.3cm}
\multicolumn{3}{c}{$z$ band (FLWO 1.2-m)} \\
\noalign{\vskip 0.3cm}
 2454223.741853\dotfill  &      0.99446   &     0.00160   \\
 2454223.745823\dotfill  &      0.99218   &     0.00160   \\
 2454223.746321\dotfill  &      0.99209   &     0.00160   \\
\noalign{\vskip 0.3cm}
\multicolumn{3}{c}{$z$ band (FLWO 1.2-m)} \\
\noalign{\vskip 0.3cm}
 2454230.705029\dotfill  &      0.99892  &      0.00170  \\
 2454230.705550\dotfill  &      1.00057  &      0.00170  \\
 2454230.706059\dotfill  &      0.99974  &      0.00170   \\
\noalign{\vskip 0.3cm}
\multicolumn{3}{c}{$B$ band (Lowell 0.8-m)} \\
\noalign{\vskip 0.3cm}
 2454230.770620\dotfill  &      0.99944 &       0.00150  \\
 2454230.773240\dotfill  &      0.99791 &       0.00150  \\
 2454230.774280\dotfill  &      0.99852 &       0.00150   \\
\enddata 

\tablecomments{The time stamps represent the Heliocentric Julian Date
  at the time of mid-exposure. The data have been corrected for
  residual extinction effects, and the uncertainties have been
  rescaled as described in \S~\ref{sec:obs}. We intend for this table to appear in its
  entirety in the electronic version of the journal. A portion is
  shown here to illustrate its format. The data are also available
  from the authors upon request.}

\end{deluxetable}

\clearpage

\begin{deluxetable}{lcc}
\tabletypesize{\normalsize}
\tablecaption{Quadratic LD coefficients adopted for \mbox{TrES-3} and \mbox{TrES-4}\label{tbl:ldcoeff}}
\tablewidth{0pt}

\tablehead{
\colhead{Filter} & \colhead{Linear coefficient $u_1$} & \colhead{Quadratic coefficient $u_2$}
}

\startdata
\noalign{\vskip 0.3cm}
\multicolumn{3}{c}{TrES-3} \\
\noalign{\vskip 0.3cm}
$B$ & 0.6379 & 0.1792 \\
$V$ & 0.4378 & 0.2933 \\
$g$ & 0.5535 & 0.2351 \\ 
$r$ & 0.3643 & 0.3178 \\
$i$ & 0.2777 & 0.3191 \\
$z$ & 0.2179 & 0.3162 \\
\noalign{\vskip 0.3cm}
\multicolumn{3}{c}{TrES-4} \\
\noalign{\vskip 0.3cm}
$z$ &  0.1483 & 0.3600 \\
$B$ &  0.5377 & 0.2579 \\
\enddata 

\tablecomments{The assumed limb-darkening law was $I_\mu/I_0 = 1-u_1(1-\mu)-u_2(1-\mu)^2$.}

\end{deluxetable}

\clearpage

\begin{deluxetable}{lcccc}
\tabletypesize{\normalsize}
\tablecaption{Mid-transit times of \mbox{TrES-3} and \mbox{TrES-4}\label{midtrans}}
\tablewidth{0pt}

\tablehead{
\colhead{HJD} & \colhead{Uncertainty (days)} & \colhead{Epoch $E$}
}
\startdata
\noalign{\vskip 0.3cm}
\multicolumn{3}{c}{TrES-3} \\
\noalign{\vskip 0.3cm}
2454185.910430   &    0.000198 & 0    \\
2454198.973147   &    0.000223 & 10   \\
2454214.646298   &    0.000280 & 22   \\
2454215.952080   &    0.000214 & 23   \\
2454535.968246   &    0.000166 & 268  \\
2454552.948971   &    0.000147 & 281  \\
2454569.929089   &    0.000153 & 294  \\ 
2454594.745943   &    0.000253 & 313  \\
\noalign{\vskip 0.3cm}
\multicolumn{3}{c}{TrES-4} \\
\noalign{\vskip 0.3cm}
2454223.797215   &    0.000847 &  0  \\ 
2454230.904913   &    0.000656 &  7  \\
2454230.905624   &    0.001106 &  7  \\
\enddata 
\end{deluxetable}

\clearpage

\begin{deluxetable}{lc}
\tablecaption{Properties of the \mbox{TrES-3} parent star\label{star_tres3}}
\tablewidth{0pt} 
\tablehead{\colhead{~~~~~~~~~Parameter~~~~~~~~~} & \colhead{Value}}
\startdata
$T_\mathrm{eff}$ (K)\tablenotemark{a}\dotfill    & 5650~$\pm$~75\phn\phn\\
$\log g$\tablenotemark{a}\dotfill                & 4.4~$\pm$~0.1\\
$\log g$\tablenotemark{b}\dotfill                & $4.568_{-0.014}^{+0.009}$\\
$v \sin i$ (km s$^{-1}$)\tablenotemark{a}\dotfill & 1.5~$\pm$~1.0 \\
$\xi_t$ (km s$^{-1}$)\tablenotemark{a}\dotfill   & 0.85~$\pm$~0.1\phn \\
$[$Fe/H$]$\tablenotemark{a}\dotfill              & $-$0.19~$\pm$~0.08\phs \\
$\langle\log R^\prime_{HK}\rangle$\tablenotemark{a}\dotfill  & $-4.54 \pm 0.13$\phs \\
$\log\epsilon{\rm (Li)}$\tablenotemark{a}\dotfill & $< 1.0$ \\
$\rho_\star$ (g cm$^{-3}$)\tablenotemark{c}\dotfill  & $2.304 \pm 0.066$ \\
$M_\star$ ($M_\sun$)\tablenotemark{b}\dotfill    &  $0.928_{-0.048}^{+0.028}$ \\
$R_\star$ ($R_\sun$)\tablenotemark{b}\dotfill    &  $0.829_{-0.022}^{+0.015}$ \\
Age (Gyr)\tablenotemark{b}\dotfill               &  $0.9_{-0.8}^{+2.8}$ \\
$L_\star$ ($L_\sun$)\tablenotemark{b}\dotfill    &  $0.625_{-0.058}^{+0.066}$ \\
$M_V$ (mag)\tablenotemark{b}\dotfill             &  $5.39 \pm 0.11$ \\
Distance (pc)\tablenotemark{b}\dotfill           &   $228 \pm 12$\phn \\
$U$, $V$, $W$ (km s$^{-1}$)\tablenotemark{b}\dotfill  & [+27.3, +6.7, +33.0]   \\
\enddata

\tablenotetext{a}{Determined spectroscopically.}

\tablenotetext{b}{Inferred from stellar evolution models using
observational constraints (see text).}

\tablenotetext{c}{Derived observationally.}

\tablecomments{The value adopted for the solar abundance of iron is
$\log(N_\mathrm{Fe}/N_\mathrm{H})_\odot = 7.52$}
\end{deluxetable}

\clearpage

\begin{deluxetable}{lc}
\tablecaption{Properties of the \mbox{TrES-4} parent star\label{star_tres4}}
\tablewidth{0pt} 
\tablehead{\colhead{~~~~~~~~~Parameter~~~~~~~~~} & \colhead{Value}}
\startdata
$T_\mathrm{eff}$ (K)\tablenotemark{a}\dotfill    & 6200~$\pm$~75\phn\phn\\
$\log g$\tablenotemark{a}\dotfill                & 4.0~$\pm$~0.1\\
$\log g$\tablenotemark{b}\dotfill                & $4.053_{-0.042}^{+0.030}$ \\
$v \sin i$ (km s$^{-1}$)\tablenotemark{a}\dotfill & 8.5~$\pm$~0.5 \\
$\xi_t$ (km s$^{-1}$)\tablenotemark{a}\dotfill   & 1.50~$\pm$~0.05 \\
$[$Fe/H$]$\tablenotemark{a}\dotfill              & +0.14~$\pm$~0.09\phs \\
$\langle\log R^\prime_{HK}\rangle$\tablenotemark{a}\dotfill  & $-5.11 \pm 0.15$\phs \\
$\log\epsilon{\rm (Li)}$\tablenotemark{a}\dotfill & $< 1.5$ \\
$\rho_\star$ (g cm$^{-3}$)\tablenotemark{c}\dotfill  & $0.314_{-0.032}^{+0.034}$ \\
$M_\star$ ($M_\sun$)\tablenotemark{b}\dotfill    & $1.404_{-0.134}^{+0.066}$ \\
$R_\star$ ($R_\sun$)\tablenotemark{b}\dotfill    & $1.846_{-0.087}^{+0.096}$  \\
Age (Gyr)\tablenotemark{b}\dotfill               & $2.9_{-0.4}^{+1.5}$ \\
$L_\star$ ($L_\sun$)\tablenotemark{b}\dotfill    & $4.53_{-0.62}^{+0.72}$ \\
$M_V$ (mag)\tablenotemark{b}\dotfill             & $3.13 \pm 0.17$ \\
Distance (pc)\tablenotemark{b}\dotfill           & $492 \pm 39$\phn \\
$U$, $V$, $W$ (km s$^{-1}$)\tablenotemark{b}\dotfill  & [$-$43.9, $-$39.1, $-$6.9]   \\
\enddata

\tablenotetext{a}{Determined spectroscopically.}

\tablenotetext{b}{Inferred from stellar evolution models using
observational constraints (see text).}

\tablenotetext{c}{Derived observationally.}

\tablecomments{The value adopted for the solar abundance of iron is
$\log(N_\mathrm{Fe}/N_\mathrm{H})_\odot = 7.52$}
\end{deluxetable}

\clearpage

\begin{deluxetable}{lc}
\tablewidth{0pc}
\tablecaption{
	Revised spectroscopic orbit and light curve solution for \mbox{TrES-3}, and
	inferred planet parameters
	\label{tab:par_tres3}
}
\tablehead{
	\colhead{~~~~~~~~~~~~~~~~~Parameter~~~~~~~~~~~~~~~~~} &
	\colhead{Value}
}
\startdata
\noalign{\vskip -9pt}
\sidehead{Light curve parameters}
~~~$P$ (days)\dotfill		&  1.30618581  (fixed)\\
~~~$T_c$ (HJD)\dotfill		&2,454,185.9104 (fixed)  \\
~~~$a/\rstar$\dotfill&  $5.926\pm0.056$\\
~~~$\rpl/\rstar$\dotfill	&  $0.1655\pm0.0020$\\
~~~$b \equiv a \cos i/\rstar$\dotfill	&  $0.840\pm0.010$\\
~~~$i$ (deg)\dotfill			 		&  $81.85\pm0.16$\phn \\
\sidehead{Spectroscopic parameters}
~~~$K$ (\ms)\dotfill                	&  $369\pm11$\phn\\
~~~$\gamma_\mathrm{HIRES}$\tablenotemark{a} (\ms)\dotfill         	 & $+369.8\pm7.1$\phn\phn\phs \\
~~~$e$\dotfill                     	 & 0\,(fixed) \\
~~~$(M_p\sin i)/(M_\star+M_p)^{2/3}$ ($M_\odot$)\dotfill  & $0.001893\pm0.000058$ \\
~~~rms (\ms)\dotfill                       &  22.0 \\
\sidehead{Planet parameters}
~~~$\mpl$ ($\mjup$)\dotfill  	& $1.910_{-0.080}^{+0.075}$ \\
~~~$\rpl$ ($\rjup$\tablenotemark{b})\dotfill	 	& $1.336_{-0.037}^{+0.031}$ \\
~~~$\rho_p$ (\gcmc)\dotfill	 	& $0.994_{-0.078}^{+0.095}$ \\
~~~$a$ (AU)\dotfill                 &  $0.02282_{-0.00040}^{+0.00023}$\\
~~~$\log \gpl$ (cgs) \dotfill         & $3.425 \pm 0.019$ \\
\enddata

\tablenotetext{a}{$\gamma_\mathrm{HIRES}$ is the center-of-mass velocity for the Keck relative velocities}
\tablenotetext{b}{The equatorial radius of Jupiter at 1 bar is $\rjup =$ 71,492 km.}

\end{deluxetable}

\clearpage

\begin{deluxetable}{lc}
\tablewidth{0pc}
\tablecaption{
Spectroscopic orbit and revised light curve solution for \mbox{TrES-4}, and
inferred planet parameters	\label{tab:par_tres4}}
\tablehead{
\colhead{~~~~~~~~~~~~~~~~~Parameter~~~~~~~~~~~~~~~~~} &
\colhead{Value}}
\startdata
\noalign{\vskip -9pt}
\sidehead{Light curve parameters}
~~~$P$ (days)\dotfill		& 3.553945 (fixed)  \\
~~~$T_c$ (HJD)\dotfill		& 2,454,230.9053 (fixed) \\
~~~$a/\rstar$\dotfill					&  $5.94\pm0.21$ \\
~~~$\rpl/\rstar$\dotfill				& $0.09921\pm0.00085$  \\
~~~$b \equiv a \cos i/\rstar$\dotfill	&  $0.766\pm0.020$ \\
~~~$i$ (deg)\dotfill					& $82.59\pm0.40$\phn \\
\sidehead{Spectroscopic parameters}
~~~$K$ (\ms)\dotfill               	& $97.4\pm7.2$\phn \\
~~~$\gamma_\mathrm{HIRES}$\tablenotemark{a} (\ms)\dotfill         	& $+23.7\pm5.8$\phn\phs \\
~~~$e$\dotfill                     	& 0\, (fixed) \\
~~~$(M_p\sin i)/(M_\star+M_p)^{2/3}$ ($M_\odot$)\dotfill  & $0.000698\pm0.000052$ \\
~~~rms (\ms)\dotfill                       &  11.1 \\
\sidehead{Planet parameters}
~~~$\mpl$ ($\mjup$)\dotfill 	&  $0.925_{-0.082}^{+0.081}$ \\
~~~$\rpl$ ($\rjup$\tablenotemark{b})\dotfill		& $1.783_{-0.086}^{+0.093}$ \\
~~~$\rho_p$ (\gcmc)\dotfill		& $0.202_{-0.032}^{+0.038}$ \\
~~~$a$ (AU)\dotfill                &  $0.05105_{-0.00167}^{+0.00079}$\\
~~~$\log \gpl$ (cgs) \dotfill        & $2.858 \pm 0.046$ \\
\enddata

\tablenotetext{a}{$\gamma_\mathrm{HIRES}$ is the center-of-mass velocity for the Keck relative velocities}
\tablenotetext{b}{The equatorial radius of Jupiter at 1 bar is $\rjup =$ 71,492 km.}

\end{deluxetable}

\clearpage

\clearpage

\clearpage

\clearpage

\clearpage

\clearpage

\clearpage

\clearpage

\end{document}